\documentclass[prb,aps,amsmath,amssymb,twocolumn]{revtex4}
\usepackage{graphicx}
\usepackage{bm}
\begin{document}

\title{Non-Born effects in scattering of electrons in a  conducting strip with low concentration of impurities}
\author{N. S. Peshcherenko\footnote{e-mail: peshcherenko@itp.ac.ru}}
\affiliation{L. D. Landau Institute for Theoretical Physics, Moscow 119334, Russia}
\author{A. S. Ioselevich\footnote{e-mail: iossel@itp.ac.ru}}
\affiliation{Condensed-matter physics laboratory, National Research University Higher School of Economics, Moscow 101000, Russia,}
\affiliation{L. D. Landau Institute for Theoretical Physics, Moscow 119334, Russia}

\address{}
\date{\today}

\begin{abstract}

We extend the theory of non-Born effects in resistivity $\rho$ of clean conducting tubes (developed in our previous work \cite{IosPeshPRB2019}) to ``strips'' -- quasi-one-dimensional structures in 2D conductors. Here also an original Van Hove singularity in dependence of $\rho$ on the position of chemical potential $\varepsilon$ is asymmetrically split in two peaks for attracting impurities. However, since amplitudes of scattering at impurities depend on their positions, these peaks are inhomogeneously broadened. Strongest broadening occurs in the left peak, arising, for attracting impurities,  due to scattering at quasistationary levels. In contrast with the case of tube these levels form not a unique sharp line, but a relatively broad impurity band with a weak quasi-Van Hove  feature on its lower edge. 
Different parts of $\rho(\varepsilon)$ are dominated by different groups of impurities:  close to the minimum the most effective scatterers, paradoxically are the ``weakest'' impurities -- those, located close to nodes of the electronic wave-function, so that the bare scattering matrix elements are suppressed. The quasi-Van Hove feature at left maximum is dominated by strongest impurites, located close to antinodes.

\end{abstract}

\maketitle

\section{Introduction \label{Intro}}

The principal aim of the present work is to introduce a general frame that allows for finding a resistivity of various quasi-one dimensional systems including different types of tubes, wires and strips. Physical examples of such systems could be carbon nanotubes (both single-wall \cite{single-wall0,single-wall} and multi-wall \cite{multi-wall1,multi-wall2} ones), thin Bismouth wires \cite{Brandt77,Brandt82,Nikolaeva2008}, nanoribbons \cite{nanoribbon1,nanoribbon2}  or long constrictions in 2D semiconductor heterostructures \cite{constrictions,Thornton1986,Zheng1986} produced by gates that confine motion of 2D electrons to one dimension. The properties of all these systems may be quite different. However, we will show that at least two important classes exist -- one is topologically equivalent fo a single-wall tube and the other -- to a constriction (a strip). Although the properties of resistivity for systems within these classes are very similar, there are certain important distinctions between systems from different classes. 

\subsection{Results of the previous study: the case of tube}

In our previous studies \cite{IosPeshJETPL2018,IosPeshPRB2019} we have considered a resistivity of a clean conducting tube in a longitudinal magnetic field. From the geometrical point of view the tube was supposed to be ideally cylindrical (with a circular cross-section of radius $R$) with symmetry axis $z$. Some effects of weak geometrical disorder (e.g., fluctuations of the radius $R$) were considered earlier \cite{Ioselevich2015}. The magnetic field  plays here a role of an instrument that, due to Aharonov-Bohm effect \cite{AB59}, allows for convenient and smooth shifting of the Fermi-level position $E_F$ with respect to the transversal quantization subbands. The wave-length of electrons at the Fermi level is assumed small: 
\begin{align}
\lambda_F\ll R,   \label{crcr2}
\end{align}
so that at least their transversal motion is quasiclassical.

 Some rare impurities sitting on the surface of the tube were assumed to be weak and short-range ones, so that the electronic scattering is isotropic and can be characterized by  a single small constant -- a dimensionless  amplitude of  scattering $\lambda\ll 1$. Two-dimensional concentration of impurities $n_{\rm imp}^{(2)}$ was assumed to be low enough, so that the mean free path $l\gg R$. Note that the opposite case of strong disorder ($l\ll R$) was thourougly studied both experimentally and theoretically  within the frame of weak localization theory \cite{AAS81,AASSS82,AS87}. 

For a clean tube, Van Hove singularities  are present in the dependence of resistivity $\rho(\varepsilon)$  on the dimensionless distance $\varepsilon$ between the Fermi level and the bottom of closest one-dimensional subband \cite{VanHove}. Taking scattering into account certainly should smoothen the singularities.

As we have shown in \cite{IosPeshPRB2019}, two different regimes with respect to dimensionless impurity  concentration $n=(2\pi R)^2n_{\rm imp}^{(2)}$ are possible. Namely, there exists certain crossover concentration
\begin{align}
n_c=|\lambda|,   \label{crcr1}
\end{align}
which distinguishes two cases that we discuss below.

\subsubsection{Relatively high impurity concentration}

For $n\gg n_c$ the scattering can be adequately described within the Born approximation and the Van Hove singularities are simply rounded at $\varepsilon\sim\varepsilon_{\min}$. The width $\varepsilon_{\min}$ by the order of magnitude could be found from the condition  $\varepsilon_{\min}\sim\tau^{-1}(\varepsilon_{\min})$, where $\tau^{-1}(\varepsilon)$ is (essentially energy dependent) the Born scattering rate. 

The exact shape of the density of states $\nu(\varepsilon)$ and the resistivity $\rho(\varepsilon)$, though being well known  for strictly one-dimensional systems (see \cite{BychkovDykhne1966,FrishLloyd1960,Halperin1960,LifshitsGredeskulPastur1982}), for quasi-one-dimensional systems of interest seems to be still not well understood.  Previously it has been studied (see, e.g., \cite{KearneyButcher1987,HuegleEgger2002}) under various versions of self-consistent Born approximations \cite{self-consistent,self-consistent1,self-consistent2,Lee}. In \cite{IosPeshPRB2019} we have also introduced our own variant  of such approximation. In the present paper, however, we will revisit this problem and derive an analytic solution for quasi-one dimensional resistivity, based upon using the exact strictly-one-dimensional results \cite{FrishLloyd1960,LifshitsGredeskulPastur1982}.

\subsubsection{Low impurity concentration}

For $n\ll n_c$ the single-impurity non-Born effects in scattering become essential despite the weakness of scattering ($|\lambda|\ll 1$). The peak of the resistivity is asymmetrically split in a Fano-resonance manner (see \cite{fano,fano1}), however with a more complex structure \cite{IosPeshJETPL2018}. Namely, for $\varepsilon>0$ there is a broad maximum with $\rho_{\max}^{(+)}\sim n|\lambda|$ at $\varepsilon\sim \lambda^2$, while for $\varepsilon\sim n^2\ll \lambda^2$ there is a deep minimum with $\rho_{\min}\sim n^3$. The behaviour of $\rho$ below the Van Hove singularity (at  $\varepsilon<0$) depends on the sign of $\lambda$. In case of repulsion $\rho$ monotonically grows with $|\varepsilon|$ and saturates at $\rho_0\sim\lambda^2n$  for $|\varepsilon|\gg \lambda^2$. In case of attraction $\rho$ has sharp maximum with $\rho_{\max}^{(-)}\sim n$ at $|\varepsilon|\propto \lambda^2$. The latter feature is due to resonant scattering at quasistationary bound states that inevitably arise just below the bottom of each subband for any attracting impurity. 

\subsection{Two universal classes of quasi-one dimensional systems: tubes and strips}

In this paper we consider only those quasi-one-dimensional systems which 
(i) are made of a smoothly deformed connected piece of a two-dimensional material
(ii) are geometrically (i.e., without taking impurities into account) homogeneous along the $z$-axis. 

\begin{figure}[ht]
\includegraphics[width=1\linewidth]{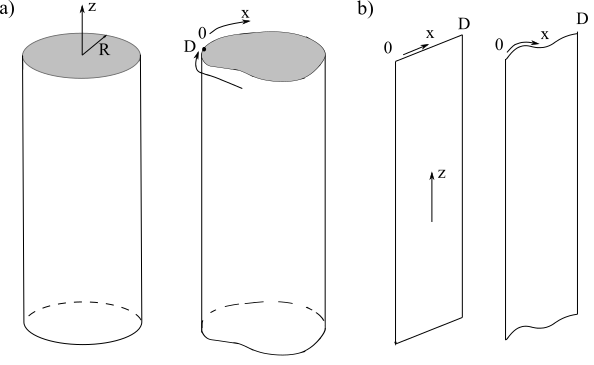}
\caption{Two topological classes of quasi-one dimensional systems. (a) tubes (cylindrical and deformed), strips (flat and deformed). The curvilinear coordinate $x$ ($0<x<D$) runs along the circumference of the cross-section normal to the axis $z$.}
\label{classes}
\end{figure}

Then we can single out two principal topological system classes: 

(a) circular or smoothly deformed cylinders (see Fig.\ref{classes}(a)) 

(b) flat or smoothly deformed strips  (see Fig.\ref{classes}(b)) 

 One can of course imagine more sophisticated topological classes with nontrivial self-crossings, but they are much less likely to be found in the nature. They also can be studied without any serious complications, if necessary. 
 
 Let us introduce curvilinear coordinates $z,x$ where the $x$-axis is perpendicular to $z$ and locally tangential to the surface (see Fig.\ref{classes}). Then, if the local radius of curvature $R(x)\gg \lambda_F$, the hamiltonian of an electron in the leading adiabatic approximation can be written as a standard two-dimensional one in these coordinates:
\begin{align}
\hat{H}_0=-\frac{\hbar^2}{2m^*}\left(\frac{\partial^2}{\partial z^2}+\frac{\partial^2}{\partial x^2}\right),\quad 0<x<D,
\\
\psi(x,z)=\exp(ikz)\chi_m(x),   \label{curvili1}
\end{align}
where $D$ is the perimeter of the cross-section. In particular, for a cylinder considered in \cite{IosPeshJETPL2018,IosPeshPRB2019}, $D=2\pi R$ while in the case of strip $D$ is its width. The equation \eqref{curvili1} is valid for both classes described above, the only difference being the boundary conditions for the transversal wave-function $\chi (x)$ at $x=0$ and at $x=D$. For the class (a) these are the periodic ones,
\begin{align}
\chi_m (0)=\chi_m(D),\quad \chi_m=\sqrt{1/D}\exp\{2\pi imx/D\}
   \label{curvili2}
\end{align}
where $m=0,\pm 1,\pm 2,\ldots$,
 while for class (b) the  zero boundary conditions apply.
 \begin{align}
\chi_m (0)=\chi_m(D)=0,\quad \chi_m=\sqrt{2/D}\sin\{\pi mx/D\}
   \label{curvili3}
\end{align}
where $m=1,2,\ldots$
Within each class the systems differ only in the length-scale $D$ what can be easily eliminated by proper choice of units.

Thus, it is enough to study only one representative for each class: say, a cylindrical tube for the class (a), and a flat strip for class (b). The first part of this task was already done in \cite{IosPeshJETPL2018,IosPeshPRB2019}, so in this paper we will mostly concentrate on the second one.

In general we assume the following conditions to be fulfilled: The condition of low concentration
 \begin{align}
n\equiv D^2n_{\rm imp}^{(2)}\ll 1,
   \label{coco1}
\end{align}
the condition of weak scattering
 \begin{align}
|\lambda|\ll 1,\quad\mbox{(weak impurities)},
   \label{coco2}
   \end{align}
   the quasiclassical condition (large number $N$ of open channels)
 \begin{align}
N\equiv D/\lambda_F \gg 1,
   \label{coco3}
\end{align}
the condition of quasi-one-dimensionality
 \begin{align}
D \ll l.
   \label{coco4}
\end{align}

\subsection{Specific physical features   of strips}

Thus, we will focus on a case of ``strip'' --  a clean conducting constriction of constant width $D$ in a two-dimensional electron gas.  We will see that there is some specifics in the physics of non-Born effects in a strip that distinguishes the strip from the cylinder. The origin of the difference is the different character of the electron eigenfunctions in a strip and  in a cylinder. While in a cylinder $|\chi_m(x)|^2={\rm const}$ depends neither on $x$, nor on $m$, in a strip $|\chi_m(x)|^2$ are essentially inhomogeneous and, therefore, the relevant scattering matrix elements depend on the position of impurity: they are suppressed for those impurities that are placed close to the nodes of the transverse wave-function $\chi_N(x)\propto\sin\{\pi Nx/D\}$ of the resonant subband $N$ and enhanced for those impurities that are close to antinodes. 

Within the Born regime (for relatively high concentration of impurities $n>n_c$) the only consequence of this fact is the replacement of the unique scattering amplitude $\lambda$ existing in the case of a tube by the averaged one. This replacement changes only some numerical factors in the final results for the resistivity and the density of states. 

It is not the case for the strongly non-Born regime ($n<n_c$). Here, as we will see, for any given $\varepsilon$ there is a certain specific group of impurities that scatter the charge carriers most effectively. In particular, for small $\varepsilon$ this group consists of ``weak'' impurities, sitting quite close to the nodes, so that the bare (Born) scattering amplitudes for them are considerably suppressed. 

Most spectacularly the specifics of scattering in a strip is manifested in case of attracting impurities ($\lambda<0$). Here, a solitary quasistationary level at $\varepsilon_{\rm qs}$, existent in a tube, is inhomogeneously broadened, forming an ``impurity band'' with relatively sharp edges. The upper edge of the impurity band  lies at $\varepsilon_{\rm qs}=0$ and for electrons with small $|\varepsilon|$ the scattering is dominated by ``weak'' impurities that have especially shallow quasistationary levels with small $\varepsilon_{\rm qs}\sim -|\varepsilon|$. At the lower edge of the impurity band a Van Hove-like feature arises in the resistivity, with the principal contribution coming from the ``strong'' impurities, sitting close to antinodes.

\section{principal results\label{principal results}}

In this Section we will summarize the main results of the paper. 

\subsection{Units and definitions\label{Units and definitions}}

Throughout the paper we will use $D$ as a unit for length and
\begin{align}
E_D=\frac{2\hbar^2\pi^2}{m^*D^2}\label{main-cond21ew}
\end{align}
as a unit for energy. As reference values for the density of states $\nu$, the scattering rate $\tau^{-1}$, and the resistivity $\rho$ we will use their values away from Van Hove singularities, directly related to the characteristics of the underlying two-dimensional material
 \begin{align}
\nu_0=\pi,\quad\frac{1}{\tau_0}=2n\left(\frac{\lambda}{\pi}\right)^{2},\quad	\rho_0=\frac{8\pi}{N^2e^2\tau_0}. \label{bound0qew1f}
\end{align}

\subsection{Two sorts of non-Born effects\label{Two sorts of non-Born effects}}

Clearly, for large enough $|\varepsilon|$ the scattering can be treated perturbatively. At low $|\varepsilon|$ nonperturbative effects show up. There are two sorts of these effects: (i) the single-impurity non-Born effects (they are due to multiple scattering at the same impurity) and (ii) multi-impurity ones.
For relatively high concentration $n_c\ll n\ll 1$ there are the following two relevant energy scales:
\begin{align}
\overline{U}=\left(\frac{n}{\pi}\right)\left(\frac{\lambda}{\pi}\right),\quad \varepsilon_{\min}^{\rm (B)}\equiv\left(\frac{n}{\pi}\right)^{2/3}\left(\frac{\lambda}{\pi}\right)^{4/3}\ll \overline{U},
\label{scales-gen0}
\end{align}
Both the multi-impurity non-Born effects and the single-impurity ones become essential below the same energy scale, at $|\varepsilon|\lesssim\varepsilon_{\min}^{\rm (B)}$.

For low concentration $n\ll n_c$ the two different scales are relevant: 
\begin{align}
\varepsilon_{\rm nB}\equiv\left(\frac{\lambda}{\pi}\right)^2,\quad \varepsilon_{\min}^{\rm (nB)}\equiv\left(\frac{n}{\pi}\right)^2\ll \varepsilon_{\rm nB},
\label{scales-gen}
\end{align}
Upon lowering of $|\varepsilon|$ first the single-impurity effects come into play at $|\varepsilon|\sim\varepsilon_{\rm (nB)}$ and only at $|\varepsilon|\sim\varepsilon_{\min}^{\rm (nB)}\ll\varepsilon_{\rm (nB)}$ they are accompanied by multi-impurity ones.

\subsection{
High concentration of impurities: exact result for quasi-one-dimensional systems\label{Born approximation: exact result for quasi-one-dimensional systems}}

For strictly one-dimensional systems  the exact results for the density of states and for the resistivity are well known (see, e.g., \cite{LifshitsGredeskulPastur1982,self-consistent2,Revew1D}). However, finding these quantities for a quasi-one-dimensional systems is, in principle, a much more sophisticated problem. 

\begin{figure}[ht]
\includegraphics[width=0.9\linewidth]{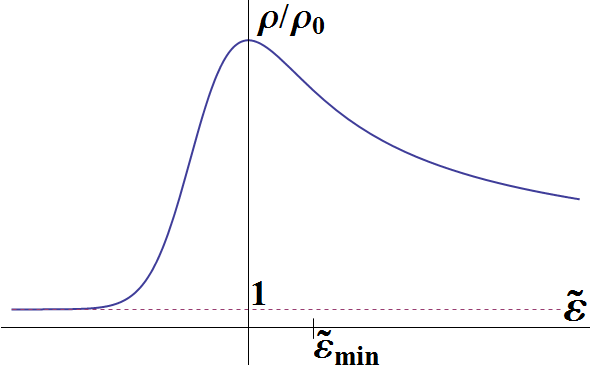}
\caption{ $\rho(\varepsilon)$ dependence near Van Hove singularity for the case $n\gg n_c=|\lambda|$. Note that all the energies are counted from the average impurity potential $\bar{U}$: $\tilde{\varepsilon}=\varepsilon-\bar{U}$. For $\tilde{\varepsilon}<0$ the density of states in the resonant subband falls off exponentially (Lifshits tail) while for $\tilde{\varepsilon}>0$ it decays, according to perturbation theory result, much slower (as $1/\sqrt{\tilde{\varepsilon}}$).}
\label{plot_res_exact}
\end{figure}

We show that under condition of high concentration of impurities  ($n_c\ll n\ll 1$) the resistivity of quasi-one-dimensional system may be expressed in terms of the exact density of states (not the exact resistivity!) of the corresponding strictly one-dimensional one. In the latter problem for $n_c\ll n\ll 1$ the random potential produced by the impurities can be reduced to the gaussian one, so that the density of states could be easily found \cite{FrishLloyd1960,Halperin1960}. As a result, we obtain
\begin{align}
\frac{\rho(\tilde{\varepsilon})}{\rho_0}=\frac{\nu(\tilde{\varepsilon})}{\nu_0}\approx1+\varepsilon_{\min}^{-1/2}Y\left(\frac{\varepsilon-\overline{U}}{\varepsilon_{\min}}\right),\label{dihy1bv}
\end{align}
 \begin{align}
\varepsilon_{\rm min}=\varepsilon_{\min}^{\rm (B)}\left\{\begin{aligned}1,& \quad(\mbox{tube}),\\
(2/3)^{-2/3},& \quad(\mbox{strip}),
\end{aligned}
\right.
\label{bobo1axx}
\end{align}
\begin{align}
Y\left(q\right)=\frac{2}{\sqrt{\pi}}\frac{\partial}{\partial q}\left(\int_0^{\infty}\frac{dx}{\sqrt{x}}\exp\left\{-xq-\frac{x^3}{12}\right\}\right)^{-1}.
\label{dihy1lw}
\end{align}
The shift $\overline{U}$ of the Van Hove singularity is nothing else but the averaged in space  potential created by impurities (positive for repulsing impurities with $\lambda>0$ and negative for attracting ones with $\lambda<0$). It is the same for both cases of tube and strip, while the width $\varepsilon_{\rm min}\sim \varepsilon_{\min}^{\rm (B)}$ differs in two cases by numerical factor $(2/3)^{-2/3}$. Note that the shift is large compared to the width of the singularity: $|\overline{U}|\gg \varepsilon_{\rm min}$.
The shape of  $\rho(\varepsilon)$ is plotted in Fig.\ref{plot_res_exact}.

\subsection{Non-Born effects for low concentration of impurities: strip vs tube\label{Non-Born effects: specifics of strip}}

As we have shown in \cite{IosPeshJETPL2018,IosPeshPRB2019} for the case of tube, in the range of low impurity concentration $n\ll n_c$ and relatively low energies $|\varepsilon|\lesssim\varepsilon_{\rm nB}$ the non-Born effects in scattering lead to strong energy-dependent renormalization of the scattering amplitudes: $\lambda\to \tilde{\Lambda}^{{\rm (ren)}}$. In the intermediate range $\varepsilon_{\min}^{\rm (nB)}\ll |\varepsilon|\lesssim\varepsilon_{\rm nB}$
the renormalization effects remain single-impurity ones so that the process may still be described in terms of the scattering amplitude at each individual impurity: $\lambda\to \tilde{\Lambda}^{{\rm (ren)}}$. We show that the same is true also for the case of a strip, but the renormalization here depends on the position of the impurity:
\begin{align}
\tilde{\Lambda}^{{\rm (ren)}}_i=\frac{\sqrt{-\varepsilon/\varepsilon_{\rm nB}}|\lambda|(1-i\lambda)}{{\rm sign}\lambda\sqrt{-\varepsilon/\varepsilon_{\rm nB}}-(1-i\lambda)2t_i},
\label{nonres_scat_tube5as2}
\end{align}
where we have introduced 
\begin{align}
2t_i\equiv |\chi_{N}( x_i)|^2=\left\{\begin{aligned}1,& \quad(\mbox{tube}),\\
1-\cos (2\pi N x_i),& \quad(\mbox{strip}).
\end{aligned}
\right.
\label{nonres_scat_tube6ewre}
\end{align}
For a long enough system the scattering amplitude enters the resistivity being effectively averaged over the positions of impurities:
\begin{align}
\frac{\rho(\varepsilon)}{\rho_0}=\frac{\tau_0}{\tau_{\rm nonres}(\varepsilon)}=-\frac{1}{\lambda^2}\int_0^1 d x\,{\rm Im}\;\{\tilde{\Lambda}^{\rm (ren)}( x,\varepsilon)\}.
\label{res_repuls02e}
\end{align}

\subsection{Non-Born resistivity: repulsive impurities\label{Non-Born effects: repulsive impurities}}

\begin{figure}[ht]
\includegraphics[width=0.9\linewidth]{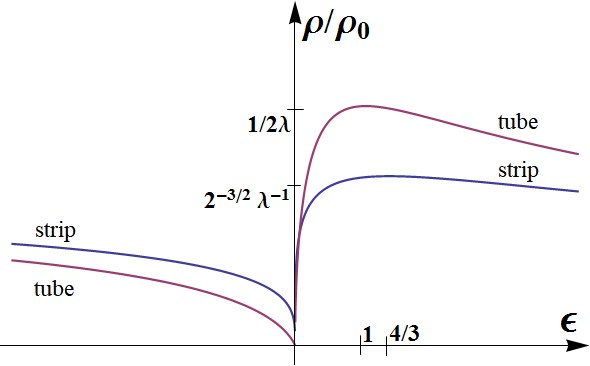}
\caption{ The dependence of resistivity $\rho$ on $\epsilon\equiv\varepsilon/\varepsilon_{\rm nB}$   for both cases of tube and strip (only repulsing impurities, $\lambda>0$).  Note that the maximum for the strip (at $\epsilon=4/3$) is lower and broader  than that for the  tube (at $\epsilon=1$). This is an inhomogeneous broadening due to the dependence of bare scattering amplitude on the position of impurity.}
\label{nonBornpos}
\end{figure}

For repulsive impurities the averaging gives the following asymptotic behavior: 
\begin{align}
\frac{\rho(\varepsilon)}{\rho_{0}}\approx\frac{1}{|\lambda|}\;\left\{
	\begin{aligned}
	\frac12(\varepsilon/\varepsilon_{\rm nB})^{1/4},\quad	&\mbox{$\varepsilon_{\min}^{\rm (nB)}\ll\varepsilon\ll \varepsilon_{\rm nB}$},
\\(\varepsilon/\varepsilon_{\rm nB})^{-1/2},\quad  & \mbox{$\varepsilon\gg \varepsilon_{\rm nB}$},
	\end{aligned}\right.\quad (\varepsilon>0),
\label{rho_rep_energy+1}		
\end{align}
\begin{align}
\frac{\rho(\varepsilon)}{\rho_{0}}\approx\left\{
	\begin{aligned}
\frac{1}{2\sqrt{2}}|\varepsilon/\varepsilon_{\rm nB}|^{1/4},\quad  & \varepsilon_{\min}^{\rm (nB)}\ll|\varepsilon|\ll\varepsilon_{\rm nB},\\
1,\quad	&|\varepsilon|\gg\varepsilon_{\rm nB},
	\end{aligned}\right.\quad (\varepsilon<0),
\label{rho_rep_energy-1}		
\end{align}
Analytic formulas for arbitrary $\varepsilon/\varepsilon_{\rm nB}$ are given in Section \ref{Non-Born resistivity: repulsing impurities}; they are also plotted in Fig.\ref{nonBornpos} together with the similar results obtained for a tube.

It is interesting that for $|\varepsilon|\ll\varepsilon_{\rm nB}$ the scattering is dominated by weak impurities with
\begin{align}
|\chi_{N}( x_i)|^2\sim\sqrt{|\varepsilon|/\varepsilon_{\rm nB}}
\label{parad1}
\end{align}
This paradoxical enhancement of their role is explained by the resonant  scattering at virtual levels arising on the unphysical sheet of complex energy (see \cite{opticaltheorem}). Note also that the resonant  character of low-energy scattering  leads to slower ($\rho\propto |\varepsilon|^{1/4}$) decreasing of resistivity at $|\varepsilon|\to 0$, as compared to the case of tube, where $\rho\propto |\varepsilon|^{1/2}$.
 
\subsection{Non-Born resistivity: attractive impurities. Distinguished role of quasistationary states.\label{Non-Born effects: attractive impurities}}

Above the Van Hove singularity, for $\varepsilon>0$ the resistivity depends only on $\lambda^2$ and the result \eqref{rho_rep_energy+1} is valid for attracting impurities as well. Below the Van Hove singularity, due to the presence of quasistationary states, for the attracting case there are two distinct ranges of energy:

Outside the impurity band ($\varepsilon<-4\varepsilon_{\rm nB}$):
\begin{align}
\frac{\rho(\varepsilon)}{\rho_{0}}\approx\left\{\begin{aligned}8\sqrt{2}\left(\frac{|\Delta\varepsilon|}{\varepsilon_{\rm nB}}\right)^{-3/2}, & \quad |\lambda|\ll\frac{|\Delta\varepsilon|}{\varepsilon_{\rm nB}}\ll 1,\\
1,& \quad |\varepsilon|\gg \varepsilon_{\rm nB}.
\end{aligned}
\right.
 \label{deni2pp4s99t}
\end{align}
where $\Delta\varepsilon\equiv\varepsilon+4\varepsilon_{\rm nB}<0$.
Within the impurity band ($-4\varepsilon_{\rm nB}<\varepsilon<0$):
\begin{align}
\frac{\rho(\varepsilon)}{\rho_{0}}
\approx\frac{1}{|\lambda|\sqrt{2}}\left\{
	\begin{aligned}
|\varepsilon/\varepsilon_{\rm nB}|^{1/4},\quad  & \varepsilon_{\min}^{\rm (nB)}\ll|\varepsilon|\ll \varepsilon_{\rm nB},\\
4\left(\frac{\Delta\varepsilon}{\varepsilon_{\rm nB}}\right)^{-1/2},\quad	&|\lambda|\ll\frac{\Delta\varepsilon}{\varepsilon_{\rm nB}}\ll 1.
	\end{aligned}\right.\label{deni2pp4po33wst}
\end{align}
For any given energy in this range the resistivity is dominated by scattering at resonant impurities with such $ x_i$ that $\varepsilon_{\rm qs}( x_i)\approx \varepsilon$. In particular, it means that close to the upper edge of the impurity band (at $\varepsilon\to 0$), the main contribution to the resistivity comes from weak impurities that satisfy the condition \eqref{parad1}. In contrast with the repulsive case, the resonant scattering here is provided not by virtual, but by quasistationary states.

  Close to the  lower edge, at $\varepsilon\to-4\varepsilon_{\rm nB}$, a two-side Van Hove-like singularity arises, divergent as $|\Delta\varepsilon|^{-1/2}$ from the side of the impurity band and as $|\Delta\varepsilon|^{-3/2}$ from the opposite side. The width of this singularity is $\Gamma\approx 8|\lambda|\varepsilon_{\rm nB}$. Close to this singularity the main contribution to resistivity comes  from strong impurities with maximal possible $|\chi_{N}( x_i)|^2\approx 2$. 
    
\begin{figure}[ht]
\includegraphics[width=0.9\linewidth]{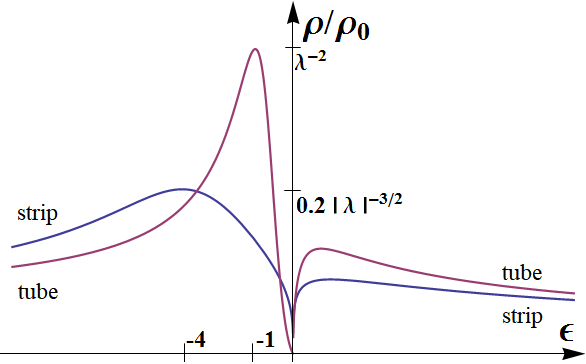}
\caption{The same as in Fig.\ref{nonBornpos} for attracting impurities, $\lambda<0$.  Note that for the strip the left maximum (at $\epsilon=-4$) is also strongly  broadened  compared to that of tube (at $\epsilon=-1$). }
\label{nonBornneg}
\end{figure}

 These results are plotted  in Fig. \ref{nonBornneg}. More detailed analytical results can be found in Section \ref{attracting impurities0}.

\subsection{The central dip in resistivity\label{The central dip in resistivity}}

\begin{figure}[ht]
\includegraphics[width=0.9\linewidth]{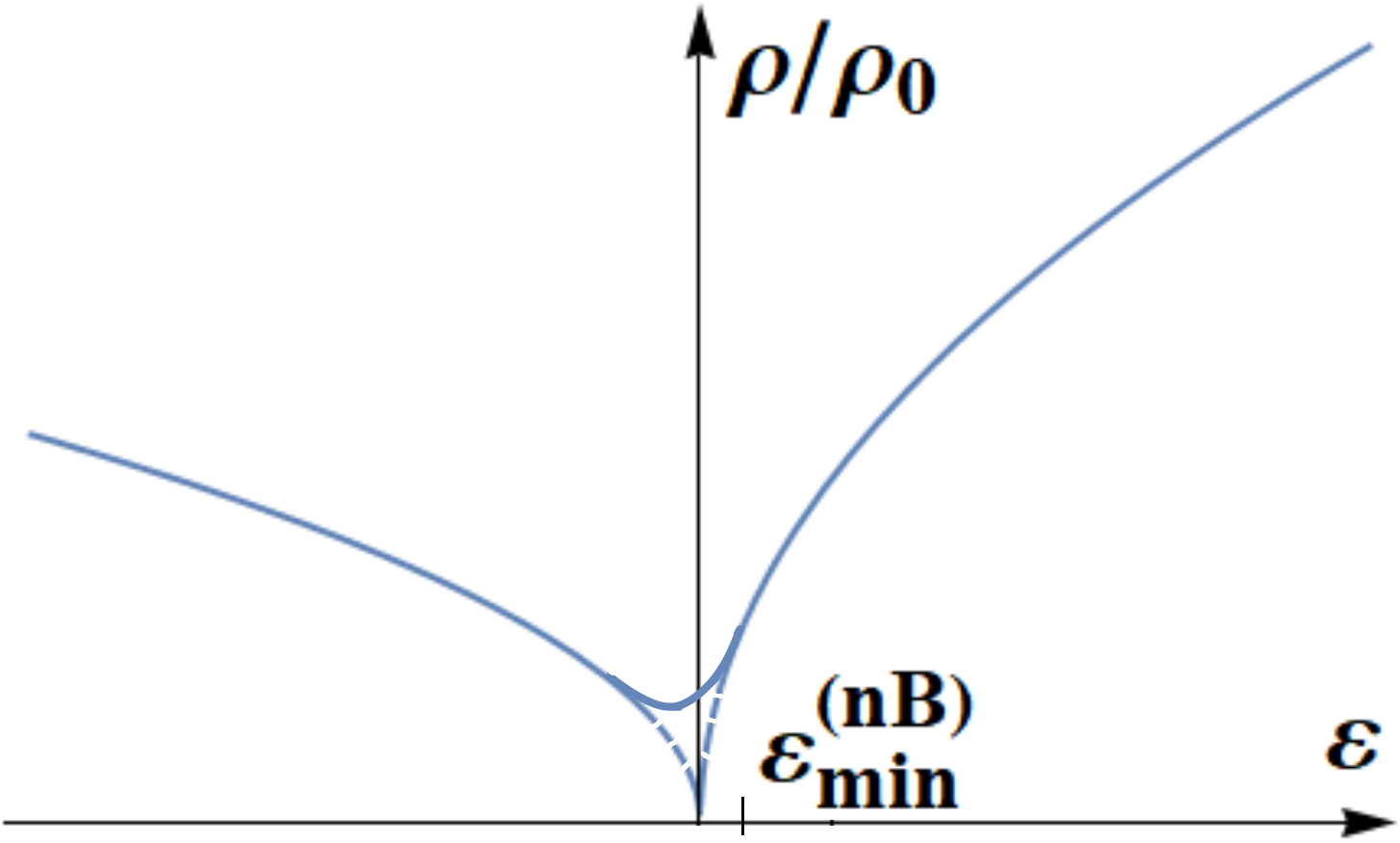}
\caption{Qualitative sketch of behaviour of the resistivity  in the range $|\varepsilon|\lesssim\varepsilon_{\min}^{\rm (nB)}$ where the single impurity approximation breaks down: $\rho(\varepsilon)$ does not go ultimately to zero, but saturates at some small but finite value.} 
\label{dip}
\end{figure}

The above results were obtained under the condition $\varepsilon_{\min}^{\rm (nB)}\ll|\varepsilon|$. Inside the range $|\varepsilon|\lesssim\varepsilon_{\min}^{\rm (nB)}$ the single-impurity approximation fails and the coherent interference of scattering at different impurities becomes essential. In this article we do not discuss the corresponding physics, though it appears to be quite tractable, again with the aid of exact solutions for the strictly one-dimensional problem. Such discussion will be given in a separate publication. Here we only want to stress that in reality, in contrast with Figs.\ref{nonBornpos}, \ref{nonBornneg}, the resistivity of course does not exactly vanish at $|\varepsilon|\to 0$, but remains finite reaching a deep minimum at some
$|\varepsilon|\sim\varepsilon_{\min}^{\rm (nB)}$ (see Fig. \ref{dip}).

\section{Structure of the paper\label{Structure of the paper}}

Our paper is organized as follows: In section \ref{An ideal strip} we remind well known facts from quantum mechanics of an electron living on a strip. In section \ref{The Born approximation1} we review the Born scattering at point-like impurities in a strip (scattering rates in subsection \ref{The scattering rates} and the conductivity in subsection \ref{The conductivity}). In section \ref{smearBorn1} we qualitatively discuss possible limitations for the Born approximation and corresponding mechanisms of smearing of Van Hove singularity. 

Section \ref{The Born approximation: exact results} is devoted to exact approach to the smearing of singularity, relevant in the case of ``high ''concentration of impurities: $n_c\ll n\ll 1$. In particular, in subsection \ref{A link to strictly one-dimensional systems} we establish a link between the resistivity of quasi-one-dimensional system and the density of states of a corresponding strictly one-dimensional one, and in subsection \ref{A summary of general features of resistivity in the Born case} we summarize the results, obtained for the case of high concentration. 

In section \ref{non-Born general} we switch to the case of low concentration $n\ll n_c$ and discuss general single-impurity non-Born effects and corresponding renormalization of scattering amplitudes. Then, in section \ref{Non-Born scattering rate and resistivity: general results} we  derive expressions for scattering rates and resistivity, and in sections \ref{Non-Born resistivity: repulsing impurities},\ref{attracting impurities0} we apply the obtained results to the cases of repulsing and attracting impurities, correspondingly. In particular, the subsection \ref{Attracting impurities1} deals with the impurity band, formed by quasistationary states. 

While the previous sections were dealing with the scattering rates of nonresonant states (which are the only current-carrying ones) in the  section \ref{Decay rates for resonant states and Breakdown of single-impurity approximation} we evaluate the scattering rate for resonant states, which is relevant for establishing the applicability range of our approach. The section \ref{Discussion and conclusion} is the conclusion.

\section{An ideal strip\label{An ideal strip}} 

\begin{figure}[ht]
\includegraphics[width=0.6\linewidth]{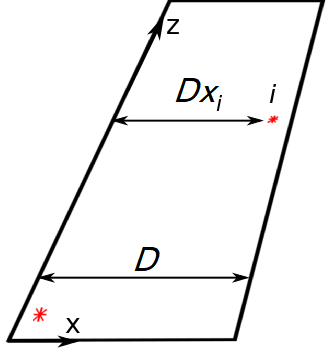}
\caption{Conducting strip of width $D$. Impurities are shown as stars. 2D electron gas lives between the edges of the strip.}
\label{setup}
\end{figure}

The eigenfunctions of electrons in an ideal strip of width $D$ are given by \eqref{curvili1} and \eqref{curvili3}, while their spectrum is
\begin{align}
    	E_{mk}=\frac{\hbar^2k^2}{2m^*}+E_m, \quad E_m=\frac{E_D}{4}m^2,	\label{main-cond21}
\end{align}
where $E_D$ is given by \eqref{main-cond21ew} and $z$ is the coordinate along the strip and $0<x<D$ is the distance from one of the strip's edges (see Fig.\ref{setup}).

\begin{figure}[ht]
\includegraphics[width=0.9\linewidth]{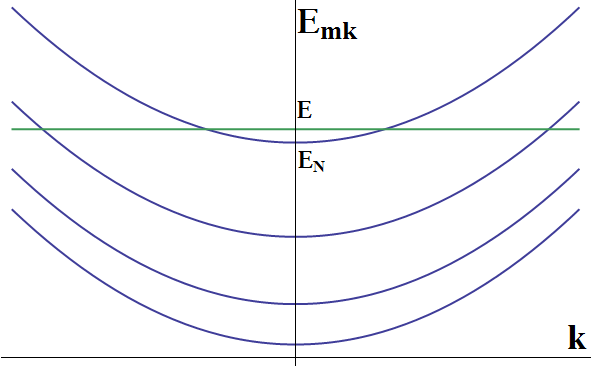}
\caption{Spectrum of an electron in an ideal strip. Subbands of the transverse quantization are shown. The Fermi level $E$ crosses all the subbands with $m\leq N$.}
\label{subbands}
\end{figure}

Integer $m$ is the transverse quantum number, $k$ is the momentum along the strip, and  $E_m$ has the meaning of position of the bottom of $m$-th one-dimensional subband. A schematic picture of the subbands is shown in Fig. 
\ref{subbands}.

  The density of states in each subband is
\begin{align}
   \nu_m(E)=\int \frac{dk}{2\pi}
   \delta\left(E-E_m-\frac{k^2}{2m^*}\right)=\nonumber\\=\frac{2}{2\pi}\sqrt{\frac{m^*}{2(E-E_m)}}\theta(E-E_m).
   \label{main-cond2rs1}
\end{align}
We will measure all energies in the units of $E_D$ and all distances in the units of $D$:
\begin{align}
x\to D x,\quad   E-E_m\to E_D\varepsilon_m.
   \label{unitsw1}
\end{align}
For brevity,  we will introduce
\begin{align}
\varepsilon\equiv\varepsilon_N
   \label{unitsw1u}
\end{align}
with $N$ being the label of the subband closest to the Fermi level. The partial densities of states in the dimensionless variables
\begin{align}
\nu_m(E)\equiv\frac{\nu_m(\varepsilon)}{DE_D},\quad\nu_m(\varepsilon)=\frac{\theta(\varepsilon_m)}{\sqrt{\varepsilon_m}}.
   \label{unitsw1p}
\end{align}

We are interested in semiclassical case when $E_D\ll E$ or $\varepsilon_0\equiv E/E_D\gg 1$. Under this condition the label $N$ of the resonant state (which is the same as the number of open channels in the system) is large:
\begin{align}
N\approx 2\sqrt{\varepsilon_0}\gg 1.   \label{unitsw1k}
\end{align}

Then, in the leading semiclassical approximation the total density of states 
\begin{align}
   \nu(\varepsilon)=\sum_{m=1}^\infty\nu_m(\varepsilon)\approx\nu_0=\int_0^{\varepsilon_0}\frac{d\varepsilon_{m}}{\sqrt{\varepsilon_m(\varepsilon_0-\varepsilon_m)}}=\pi.
   \label{deni1y}
\end{align}
This result is valid for all $\varepsilon$ except narrow interval in the vicinity of $\varepsilon=0$ point.
In the entire range of variation of $\varepsilon$ one can write
\begin{align}
   \nu(\varepsilon)\approx\nu_0\left(1+\frac{\theta(\varepsilon)}{\pi\sqrt{\varepsilon}}\right).
   \label{deniww}
\end{align}

\section{Scattering at point-like impurities: The Born approximation\label{The Born approximation1}}

Now let us find the scattering rate and the resistivity within the lowest order in impurity potential, i.e. within Born approximation. The hamiltonian of the system reads
\begin{align}
H=H_0+V\sum_{i}\delta({\bf r} - {\bf r}_i),\quad H_0=-\nabla^2/2m^{*},
\end{align}
the  positions ${\bf r}_i$ of impurities  being randomly distributed over the surface of the strip according to the Poisson distribution with average 2D density $n_{\rm imp}^{(2)}$. The constant $V$ is related to the dimensionless scattering amplitude by
 \begin{align}
\lambda= 	m^*V/2,\qquad |\lambda|\ll 1.
	\label{bound0qq}
\end{align}
In the case of strip the scattering matrix elements depend both on the quantum numbers of scattering states and on the position of the impurity ${\bf r}_i$:
\begin{align}
V_{kk'mm'}^{(i)}=V_{kk'mm'}( x_i,z_i)=\nonumber\\=V\exp\{i(k-k')z_i\}\chi_m( x_i)\chi_{m'}( x_i),
\label{self-energyqsw1}
\end{align}
where $z_i$ and $ x_i$ characterize the position of $i$-th impurity. 

\begin{figure}[ht]
\includegraphics[width=0.8\linewidth]{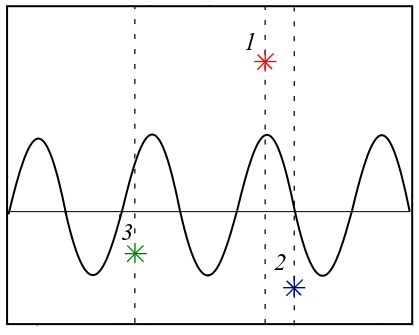}
\caption{Different groups of impurities with respect to their positions in the strip: (1) -- red star -- ``strong impurity'' (close to an antinode of the transverse wave-function $\sin (\pi N x_i)$ shown), (2) -- blue star -- ``weak impurity'' (close to a node), and (3) -- green star -- typical impurity (close neither to nodes, nor to antinodes).}
\label{oscillations}
\end{figure}

As we will see in what follows, the most important scattering processes are those in which both initial and final states belong to the resonant band: $m=m'=N$. Different impurities have different effectiveness with respect to such processes. While typical impurities are sitting in some general positions, so that $N x_i$ is close neither to integer, nor to half-integer number, there are two special groups of impurities (see Fig.\ref{oscillations}): 
\begin{enumerate}
\item {\it Weak impurities}, sitting close to nodes of the resonant transverse wave-function ($N x_i$ is close to  integer). The scattering at such impurities is suppressed.
\item {\it Strong impurities}, sitting close to antinodes ($N x_i$ is close to half-integer): these are scattering most effectively.
\end{enumerate}
We will see that these two groups  may play distinguished role and give leading contribution to the resistivity in certain ranges of parameters.

        \subsection{The scattering rates\label{The scattering rates}}

The averaged over the positions of impurities decay rate for a general state $m$:
\begin{widetext}
\begin{align}
\frac{1}{\tau_m(\varepsilon)}=n\left\langle\int dk'\sum_{m'}|V_{kk'mm'}( x)|^2\delta(E_{m'k'}-E_{mk})\right\rangle_{ x}=\frac{1}{\tau_0}\int_0^1d x\sum_{m'}
|\chi_m( x)\chi_{m'}( x)|^2
\frac{\theta(\varepsilon_{m'})}{\pi\sqrt{\varepsilon_{m'}}} =\nonumber\\=\frac{1}{\nu_0\tau_0}\left(\sum_{m'}\nu_{m'}+\frac12\nu_m\right)\approx\frac{1}{\tau_0}\left(1+\frac{\theta(\varepsilon)}{\pi\sqrt{\varepsilon}}\left\{\begin{aligned} 1,\quad& \mbox{for $m\neq N$},\\
3/2,\quad & \mbox{for $m=N$}.
\end{aligned}\right\}\right),
\nonumber
\end{align}
where we have used
\begin{align}
\int_0^1d x(2\sin^2(\pi m x))( 2\sin^2\pi m' x))=\left\{\begin{aligned} 1,\quad& m'\neq m,\\
3/2,\quad & m'=m.
\end{aligned}\right.\nonumber
\end{align}
\end{widetext}
and the fact that for large $N\gg 1$ each individual non-resonant contribution to the sum is relatively small, while the resonant one may be large, provided $\varepsilon_N\equiv\varepsilon\ll 1$.  
So, we have shown that within the Born approximation  the scattering rate is the same  for all the nonresonant states:
\begin{align}
    \frac{\tau_0}{\tau_{m\neq N}(\varepsilon)}= \frac{\tau_0}{\tau_{\rm nonres}(\varepsilon)}=\frac{\nu(\varepsilon)}{\nu_0}
    \approx  1+\frac{\theta(\varepsilon)}{\pi\sqrt{\varepsilon}},\label{deni2ppl}
\end{align}
while for the resonant state 
\begin{align}
\frac{\tau_0}{\tau_N(\varepsilon)}= \frac{\tau_0}{\tau_{\rm res}(\varepsilon)}\approx1+\frac{3\theta(\varepsilon)}{2\pi\sqrt{\varepsilon}},  \label{deni2pp2}
\end{align}

\subsection{Current-carrying states and the conductivity\label{The conductivity}}

To evaluate the conductivity of the strip (per one spin projection) we can use the Kubo formula:
\begin{widetext}
\begin{align}
\sigma=\frac{e^2}{2\pi}{\rm Tr}[\hat{v}_z\hat{G}^R\hat{v}_z\hat{G}^A]=\frac{e^2}{2\pi}\int\frac{dk}{2\pi}\sum_{m}\frac{(v^z_{km})^2}{(\varepsilon-E_{km})^2+1/4\tau_{m}^2(\varepsilon)}\approx  e^2\int\frac{dk}{2\pi}\sum_{m}(v^z_{km})^2\delta(\varepsilon-E_{km})\tau_{m}(\varepsilon),
\label{Kubo}
\end{align}
\end{widetext}
 From \eqref{Kubo} we immediately see that only non-resonant states are expected to be current carrying: the resonant state contribution to the current is suppressed by the factor $(v^{z }_{N})^2\propto \varepsilon\ll 1$. Hence, we can write
 \begin{align}
\sigma\approx  e^2D(\varepsilon)\nu_{\rm tr}(\varepsilon),\quad D(\varepsilon)=\frac12v_F^2\tau_{\rm nonres}=D_0\frac{\nu_0}{\nu(\varepsilon)}
\label{Kubo1}
\end{align}
where $D_0=\frac12v_F^2\tau_{0}$ is the two-dimensional diffusion coefficient, and  
  \begin{align}
\nu_{\rm tr}(\varepsilon)=2\int\frac{dk}{2\pi}\sum_{m}\left(\frac{v^z_{km}}{v_F}\right)^2\delta(\varepsilon-E_{km}),
\label{Kubo2}
\end{align}
is the ``transport density of states''. In contrast with the standard density of states, the transport one does not exhibit any Van Hove singularity at $\varepsilon\to 0$: the latter is suppressed by the factor $\left(v^z_{km}/v_F\right)^2$. As a result, under the semiclassical condition $\varepsilon_0\gg 1$ we can always substitute $\nu_{\rm tr}(\varepsilon)\equiv \nu_0$, even at $\varepsilon\to 0$.

 Thus, in the Born domain for the resistivity $\rho\equiv 1/\sigma$ we get a simple result:
\begin{align}
\frac{\rho(\varepsilon)}{\rho_0}=\frac{\tau_0}{\tau_{\rm nonres}(\varepsilon)}=\frac{\nu(\varepsilon)}{\nu_0}.\label{resistiv1}
\end{align}
As we see, for $\varepsilon\to0$ the resistivity $\rho(\varepsilon)$ diverges. This divergency is nothing else but the Van Hove singularity.
So, we conclude that in the range of $\varepsilon$ where the perturbation theory is applicable (i.e., neither the single-impurity non-Born effects, nor the interference of scattering at different impurities are relevant) the resistivity of a conducting strip is described by exactly  the same formulae, as the resistivity of a conducting tube, derived in \cite{IosPeshJETPL2018, IosPeshPRB2019}. One should only replace the unique scattering time $\tau$ of the tube theory by $\tau_{\rm nonres}$ of the strip theory.

\section{Smearing of the Van Hove singularity within the Born approximation\label{smearBorn1}}

It is instructive to distinguish two groups of effects nonlinear in the scattering amplitude: 
\begin{enumerate}
\item Single-impurity non-Born effects, arising due to  more accurate (i.e. nonperturbative) treatment of individual scattering acts;
\item The multi-impurity ones, coming from the interference of scattering acts at different impurities.
\end{enumerate}

 Upon approaching the Van Hove singularity the nonlinear effects of both types become stronger. However, if the concentration of impurities is relatively high, 
 \begin{align}
n\gg n_c=|\lambda|,\label{bobo1}
\end{align}
we will show that the multi-impurity effects come into play earlier than  the non-Born single-impurity effects, so that the latter do not have a chance to show up and effectively can be neglected (the Born regime). In this section we will be dealing only with this Born regime.

\subsection{Shift of the singularity\label{Shift of the singularity}}

The strongest of the multi-impurity effects that comes into play at $\varepsilon\sim\lambda n$ is quite simple. It is just the shift of the resonant subband by the average potential of impurities:
 \begin{align}
\overline{U}=\left\langle V\sum_{i}\delta({\bf r} - {\bf r}_i)\right\rangle_{{\bf r}_i}=\frac{ \lambda n}{\pi^2},\label{bobo1a}
\end{align}
It is important to note that an introduction of this shift makes sense only under condition \eqref{bobo1}. Indeed, an effective self-averaging of the potential takes place if the electronic  wave function does not change much on the scale of an inter-impurity distance $n^{-1}$, which means $n^{-1}(m\overline{U})^{1/2}\sim(\lambda/n)^{1/2}\ll 1$. The latter condition is equivalent to \eqref{bobo1}. Thus, if \eqref{bobo1} is fulfilled, one should first of all renormalize the position of the Van Hove singularity:
 \begin{align}
\varepsilon\to\tilde{\varepsilon}=\varepsilon-\overline{U}\label{bobo1b}
\end{align}
and substitute $\tilde{\varepsilon}$ instead of $\varepsilon$ in the results of the preceding section.

\subsection{Smoothing of the singularity: qualitative description\label{Smoothing of the singularity: qualitative description}}

The next multi-impurity effect is the smearing of the singularity due to scattering. This effect becomes essential at smaller energy scale $\tilde{\varepsilon}\lesssim \varepsilon_{\min}$ where the perturbation theory breaks down. The scale $\varepsilon_{\min}$ can be extracted from the condition
\begin{align}
\tau_{\rm res}^{-1}(\varepsilon_{\min})\sim\varepsilon_{\min}\label{bobo2}
\end{align}
when the resonant state become smeared. Note that  the current carrying nonresonant states become smeared at the same scale, since, as it follows from \eqref{deni2ppl} and \eqref{deni2pp2}, 
 \begin{align}
\tau_{\rm res}^{-1}\approx (3/2)\tau_{\rm nonres}^{-1},\label{bobo1bo}
\end{align}
within the Born approximation. The divergencies of both $\tau_{\rm nonres}^{-1}(\varepsilon)$ and $\tau_{\rm res}^{-1}(\varepsilon)$ are due to the divergency of the density of final states in the scattering processes.

\section{The Born approximation: exact results\label{The Born approximation: exact results}}

Electrons with energies $|\tilde{\varepsilon}|\ll |\overline{U}|$ are effectively scattered not by individual impurities, but by fluctuations of the density of impurities. Typically such fluctuations are constituted by many impurities and, therefore, their distribution is essentially gaussian. It is important to note that these gaussian fluctuations are universal: in particular, they do not depend on the character (repulsing or attracting) of individual impurities. 

The latter is not true for rare very large non-gaussian fluctuations with $|\tilde{\varepsilon}|\gtrsim |\overline{U}|$. However, these large fluctuations are not relevant, since the corresponding part of the spectrum is likely to be dominated not by the far tail of the resonant band, but by the non-resonant ones (see below).

Combining the formulas \eqref{deni2pp2} and \eqref{bobo2} we get an estimate for the width of smeared singularity
\begin{align}
\varepsilon_{\min}\sim(n\lambda^2)^{2/3},\quad \frac{\varepsilon_{\min}}{\overline{U}} \sim\left(\frac{\lambda}{n}\right)^{1/3}\ll 1,\label{bobo1c}
\end{align}
so, indeed, under condition \eqref{bobo1} the smearing occurs on the energy scale that is much smaller than the shift of the band.

\subsection{A link to strictly one-dimensional systems\label{A link to strictly one-dimensional systems}}

For energies $|\tilde{\varepsilon}|\lesssim \varepsilon_{\min}$ plane waves $\exp(ikz)$ do not provide any good approximation for the eigenfunctions of an electron in the resonant band: they should be substituted by a set of certain nontrivial wave functions $\psi_{\alpha}(z)$, depending of concrete realization of disorder. At the same time, the plane waves remain valid eigenfunctions for electrons in the current-carrying nonresonant bands. Then, if we, as before, neglect the contribution of the resonant band to the current, the conductivity still can be written in a form \eqref{Kubo}, the only modification occurs in the expression for $\tau_m(\varepsilon)$ for $m\neq N$:
\begin{widetext}
\begin{align}
\frac{1}{\tau_m(\varepsilon)}=n\left\langle\int dk'\sum_{m'\neq N}|V_{kk'mm'}( x)|^2\delta(E_{m'k'}-E_{mk})\right\rangle_{ x}
+n\left\langle\sum_{\alpha}|V_{k\alpha mN}( x,z)|^2\delta(E_{N\alpha}-E_{mk})\right\rangle_{ x,z},\label{self-en0}
\end{align}
\begin{align}
V_{k\alpha mN}( x,z)=V\exp\{ikz\}\psi^*_{\alpha}(z)\chi_m( x)\chi_N( x).\label{self-energyqsw1l}
\end{align}
So, the second term in \eqref{self-en0} can be rewritten in terms of a density of states  $\nu_{\rm res}(\tilde{\varepsilon})$ for strictly one-dimensional system
\begin{align}
\frac{1}{\pi\tau_0}\int_0^1d x|\chi_m( x)\chi_N( x)|^2
\int dz\sum_\alpha|\psi_{\alpha}(z)|^2\delta(E_{N\alpha}-E_{mk})=\frac{\nu_{\rm res}(\tilde{\varepsilon})}{\pi\tau_0}\label{self-en01}
\end{align}
\end{widetext}
We would like to stress that in our quasi-one-dimensional problem the conductivity is expressed through the exact average density of states of a purely one-dimensional problem, which is the average of one-particle Green-function (involving two $\psi$-operators). On the other hand, it is well known that the conductivity should be expressed through the exact average two-particle Green function (four $\psi$-operators), which is a much more sophisticated object than the one-particle one. 

The explanation for this paradox is as follows. There are two distinct types of $\psi$-operators in our quasi-one-dimensional problem: $\psi_{\rm nonres}$ for electrons in non-resonant bands and $\psi_{\rm res}$ -- for electrons in the resonant band. Since in our problem the resonant band does not contribute to the current directly, each term in the conductivity should necessarily contain at least two $\psi_{\rm nonres}$-operators. Remaining two $\psi$-operators may be either both of $\psi_{\rm nonres}$ type (that leads to the first term in \eqref{self-en0}), or both of $\psi_{\rm res}$-type (the second term in \eqref{self-en0}). In this term the $\psi_{\rm res}$-operators enter through the density of final states in the scattering process. There are no terms containing four $\psi_{\rm res}$-operators since the purely one dimensional contribution to the current is strongly suppressed.

So, in the Born regime we again end up with the formula \eqref{resistiv1} relating the scattering rate of nonresonant electrons (and, therefore, the resistivity) to the total density of states
 \begin{align}
\nu(\tilde{\varepsilon})\approx\nu_{\rm nonres}(\tilde{\varepsilon})+\nu_{\rm res}(\tilde{\varepsilon}),\label{dihy1nn}
\end{align}
where $\nu_{\rm nonres}(\tilde{\varepsilon})\approx\nu_0$, while the relation $\nu_{\rm res}(\tilde{\varepsilon})=\theta(\tilde{\varepsilon})(\tilde{\varepsilon})^{-1/2}$ is true only for $|\tilde{\varepsilon}|\gg \varepsilon_{\min}$. At $|\tilde{\varepsilon}|\lesssim \varepsilon_{\min}$ one should use an exact expression for $\nu_{\rm res}(\tilde{\varepsilon})$ taken from the theory of strictly one dimensional disordered systems.

\subsection{Correction to the density of states due to hybridization of bands\label{Correction to the density of states due to hybridization of bands}}

Besides the nontrivial and strong modification of $\nu_{\rm res}(\tilde{\varepsilon})$ by disorder, there is an additional effect -- hybridization between resonant and nonresonant bands due to the presence of impurities. As we will see in the next subsection, the corresponding correction to the nonresonant density of states $\nu_{\rm nonres}$ is relatively small in the relevant range of energies and can be evaluated perturbatively:
\begin{align}
\nu_{\rm nonres}(\tilde{\varepsilon})=\nu_0+\delta\nu(\tilde{\varepsilon}),\quad \delta\nu(\tilde{\varepsilon})=\nu_0\frac{d}{d\tilde{\varepsilon}}\delta\varepsilon(\tilde{\varepsilon}),
	\label{sewyt1}
\end{align}
where $\delta\varepsilon(\tilde{\varepsilon})$ is the second order (in $V$) correction to the energy $\tilde{\varepsilon}$ of certain nonresonant state arising due to scattering
\begin{align}
\delta\varepsilon(\tilde{\varepsilon})=\frac{n\lambda^2}{\pi^4}\;\;{\rm v.p.}\int\frac{\nu(\tilde{\varepsilon}^{\prime})d\tilde{\varepsilon}^{\prime}}{\tilde{\varepsilon}-\tilde{\varepsilon}^{\prime}}.
	\label{sewyt2}
\end{align}
 For $\tilde{\varepsilon}<0$ and $|\tilde{\varepsilon}|\gg \varepsilon_{\min}^{\rm (t)}$ the principal contribution to the integral in \eqref{sewyt2} comes from the states in the resonant band with energies $\tilde{\varepsilon}^{\prime}>0$ and $\tilde{\varepsilon}^{\prime}\sim|\tilde{\varepsilon}|$, so that the correction can be estimated as
\begin{align}
\delta\nu(\tilde{\varepsilon})=\frac{n\lambda^2}{\pi^4}\int_0^{\infty}\frac{d\tilde{\varepsilon}^{\prime}}{(\tilde{\varepsilon}-\tilde{\varepsilon}^{\prime})^2\sqrt{\tilde{\varepsilon}^{\prime}}}\sim\nu_0\left(\frac{\varepsilon_{\min}}{|\tilde{\varepsilon}|}\right)^{3/2}.
	\label{sewyt3}
\end{align}
Thus, we conclude that for $|\tilde{\varepsilon}|\gg\varepsilon_{\min}$ the relative correction to the density of states is indeed small.

\subsection{Exact result: the case of a tube revisited\label{Exact result: the case of a tube revisited}}

In our previous work \cite{IosPeshPRB2019} we have studied the smearing of the resistivity peak  for the case of a tube within the self-consistent Born approximation. Now we will start from revisiting the tube case in a more accurate approach exploring the exact solutions known for the strictly one-dimensional systems. Under the condition \eqref{bobo1} the one-dimensional model with identical point-like scatterers randomly distributed on a line was exhaustively studied in \cite{LifshitsGredeskulPastur1982}. It was shown that the random potential is effectively gaussian and the density of states may be evaluated with the help of Fokker-Planck equation. As a result
\begin{align}
\nu_{\rm res}^{\rm (t)}(\tilde{\varepsilon})=\nu_0\left(\varepsilon_{\min}^{\rm (t)}\right)^{-1/2}Y\left(\tilde{\varepsilon}/ \varepsilon_{\min}^{\rm (t)}\right),\label{dihy1}
\end{align}
where the superscript ${\rm (t)}$ stands for ``tube'', and
\begin{align}
 \tilde{\varepsilon}_{\rm min}^{\rm (t)}=\left(2\pi\tau_0\right)^{-2/3}=\left(\frac{n}{\pi}\right)^{2/3}\left(\frac{\lambda}{\pi}\right)^{4/3},\label{dihy1fg}
\end{align}
\begin{align}
Y\left(q\right)=\frac{2}{\sqrt{\pi}}\frac{\partial}{\partial q}\left(\int_0^{\infty}\frac{dx}{\sqrt{x}}\exp\left\{-xq-\frac{x^3}{12}\right\}\right)^{-1}.
\label{dihy1l}
\end{align}
The asymptotics of \eqref{dihy1l} at $q>0$, $q\gg 1$,
\begin{align}
Y\left(q\right)\approx\frac{1}{\pi \sqrt{q}}\label{dihy1la}
\end{align}
corresponds to the trivial perturbative result, while the asymptotics for $q<0$, $|q|\gg 1$
\begin{align}
Y\left(q\right)\approx\frac{4|q|}{\pi}
\exp\left\{-\frac43|q|^{3/2}\right\},\label{dihy1lb}
\end{align}
describes the well-known Lifshits tail of the density of states in one-dimensional system with effectively gaussian disorder. It should be noted that \eqref{dihy1lb} is indeed only an intermediate asymptotics, valid in the range $1\ll |q|\ll (n/\lambda)^{1/3}$, where the random potential is effectively gaussian,

As it was argued in \cite{IosPeshPRB2019}, there should be certain bifurcation energy $\tilde{\varepsilon}_{\rm bi}^{\rm (t)}$, such that for all energies $\tilde{\varepsilon}_{\rm bi}^{\rm (t)}<\tilde{\varepsilon}\ll 1$ the principal contribution  to the density of states comes from the resonant subband $N$:  $\nu_{\rm nonres}(\tilde{\varepsilon})\ll\nu_{\rm res}(\tilde{\varepsilon})$. Let us demonstrate that this statement is valid also for the exact solution.

The  bifurcation point $\tilde{\varepsilon}_{\rm bi}^{\rm (t)}$ can be roughly defined as the energy, at which the contribution to the density of states coming from the resonant band  becomes equal to that of the nonresonant ones: 
\begin{align}
\nu_{\rm nonres}^{\rm (t)}(\tilde{\varepsilon}_{\rm bi}^{\rm (t)})=\nu_{\rm res}^{\rm (t)}(\tilde{\varepsilon}_{\rm bi}^{\rm (t)}).\label{dihy09k}
\end{align}
As a first step, let's suppose that 
\begin{align}
|\tilde{\varepsilon}_{\rm bi}^{\rm (t)}|\gg \varepsilon_{\min}^{\rm (t)},
\label{dihy09kasa}
\end{align}
(it will be verified  soon). Then, according to \eqref{sewyt3}, $\nu_{\rm nonres}^{\rm (t)}(\tilde{\varepsilon})$ differs from $\nu_0$ only slightly, and the condition \eqref{dihy09k} takes the form
\begin{align}
\nu_0=\nu_0\left(\varepsilon_{\min}^{\rm (t)}\right)^{-1/2}Y\left(q_{\rm bi}^{\rm (t)}\right),\\ \tilde{\varepsilon}_{\rm bi}^{\rm (t)}=\varepsilon_{\min}^{\rm (t)}q_{\rm bi}^{\rm (t)},\quad q_{\rm bi}^{\rm (t)}\approx-\left(\frac38\right)^{2/3}\ln^{2/3}\left(1/\varepsilon_{\min}^{\rm (t)}\right).\label{dihy099}
\end{align}
We want to remind here again that the result \eqref{dihy099} (as well as \eqref{dihy1lb}) is valid under condition $\varepsilon_{\min}^{\rm (t)}\ll |\tilde{\varepsilon}_{\rm bi}^{\rm (t)}|\ll\overline{U}$, which is equivalent to
\begin{align}
1\ll \ln\left(1/n\lambda^2\right)\ll\left(\frac{n}{\lambda}\right)^{1/2}.\label{dihy09o}
\end{align}
In particular, the first inequality in \eqref{dihy09o} justifies our assumption \eqref{dihy09kasa}.

So, we conclude that  the contribution of the nonresonant bands is essentially unperturbed in the relevant domain $|\tilde{\varepsilon}|>|\tilde{\varepsilon}_{\rm bi}^{\rm (t)}|$. As a result, the total density of states and the resistivity of a tube can be written as 
\begin{align}
\frac{\nu^{\rm (t)}(\tilde{\varepsilon})}{\nu_0}=\frac{\rho^{\rm (t)}(\tilde{\varepsilon})}{\rho_0}\approx1+\left(\varepsilon_{\min}^{ (t)}\right)^{-1/2}Y\left(\tilde{\varepsilon}/\varepsilon_{\min}^{\rm (t)}\right)\label{dihy1bv}
\end{align}
with high accuracy in the entire range of energies $\tilde{\varepsilon}$. This dependence is plotted in Fig.\ref{plot_res_exact}.

\subsection{Exact results: the case of a strip}

Evaluation of the density of states in the case of strip is very similar to that in the case of tube.
For the energies above the bifurcation point the density of states is dominated by the states from the resonant subband and its smearing is also controlled  by scattering processes in which both initial and final states belong to the resonant subband. It means that the smearing depends on $\tau^{\rm (s)}_{\rm res}(\varepsilon)$, but not on $\tau^{\rm (s)}_{\rm nonres}(\varepsilon)$ (the superscript ${\rm (s)}$ stands for ``strip''). In this sence the problem is very similar to that of the tube,  the only difference is an additional factor $2/3$ in the definition \eqref{deni2pp2} of $\tau^{\rm (s)}_{\rm res}$, as compared to $\tau^{\rm (t)}$. This difference, however, can be removed by the redefinition of the energy scale:
\begin{align}
\varepsilon_{\rm min}^{\rm (t)}=\left(2\pi\tau_0\right)^{-2/3}\longrightarrow
 \varepsilon_{\rm min}^{ (s)}=\left(4\pi\tau_0/3\right)^{-2/3}	\label{self-energy3ajus}
\end{align}
After the rescaling the scattering rate and the density of states can be expressed in terms of the very same function $Y(q)$, which appeared in the results for the tube (see \eqref{dihy1l}, \eqref{dihy1la}, \eqref{dihy1lb}).
It also can easily be demonstrated that, exactly as in the case of tube, the nonresonant contribution to the density of states remains equal to $\nu_0$ for all negative $\tilde{\varepsilon}$ in the range $|\tilde{\varepsilon}|>|\tilde{\varepsilon}_{\rm bi}|$. As a result
\begin{align}
\frac{\nu^{\rm (s)}(\tilde{\varepsilon})}{\nu_0}=\frac{\rho^{\rm (s)}(\tilde{\varepsilon})}{\rho_0}\approx1+\left(\varepsilon_{\min}^{\rm  (s)}\right)^{-1/2}Y\left(\tilde{\varepsilon}/\varepsilon_{\min}^{\rm (s)}\right),\label{dihy1bvr}
\end{align}
\begin{align}
 \tilde{\varepsilon}_{\rm min}^{\rm  (s)}=\left(4\pi\tau_0/3\right)^{-2/3}=\left(\frac{3n}{2\pi}\right)^{2/3}\left(\frac{\lambda}{\pi}\right)^{4/3}.\label{dihy1fgt}
\end{align}
So, the difference in the resistivities of a tube and a strip is only in different numerical factors entering characteristic energy scales $\tilde{\varepsilon}_{\rm min}^{\rm  (t)}$ and $\tilde{\varepsilon}_{\rm min}^{\rm  (s)}$

\subsection{A summary of general features of resistivity in the Born case\label{A summary of general features of resistivity in the Born case}}

\begin{figure}[ht]
\includegraphics[width=0.9\linewidth]{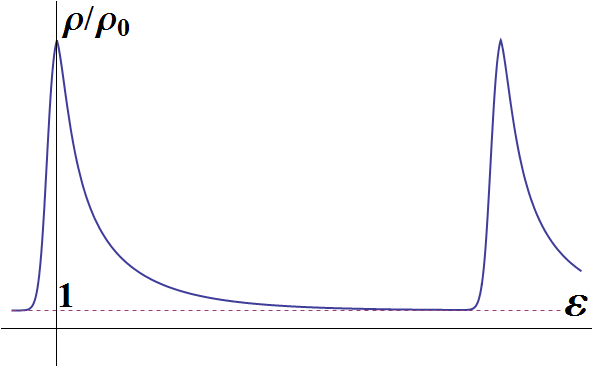}
\caption{ $\rho(\varepsilon)$ dependence for the case $n\gg n_c=|\lambda|$. The resistivity is comprised of smeared (at the scale of $\varepsilon_{\rm min}\sim (n\lambda^2)^{2/3}$) and shifted (by the value of $\bar{U}\propto n\lambda$) Van Hove singularities.}
\label{severalVH_sing}
\end{figure}

In general, the energy profile of the resistivity of a quasi-one-dimensional system with ``relatively high'' concentration (that is, for $|\lambda|\ll n\ll 1$) of weak short-range impurities consists of a set of shifted and smeared Van Hove singularities (see Fig.\ref{severalVH_sing}).

 Each individual singularity (shown in Fig.\ref{plot_res_exact}) is characterized by four distinct ranges:
\begin{enumerate}
\item Relatively smooth right slope of a shifted singularity:
\begin{align}
 \rho(\varepsilon)\approx\frac{\rho_0}{\pi(\varepsilon-\overline{U})^{1/2}},\quad \varepsilon-\overline{U}>0,\quad\frac{|\varepsilon-\overline{U}|}{\tilde{\varepsilon}_{\rm min}}\gg 1,
 \label{slo1}
\end{align}
\item Smeared core of the singularity:
\begin{align}
 \rho(\varepsilon)\sim\rho_{\max}\sim\frac{\rho_0}{\pi\tilde{\varepsilon}_{\rm min}^{1/2}},\quad \frac{|\varepsilon-\overline{U}|}{\tilde{\varepsilon}_{\rm min}}\lesssim 1,
 \label{slo2}
\end{align}
\item Exponentially steep left slope of a shifted singularity:
\begin{align}
 \rho(\varepsilon)\approx\frac{4\rho_0}{\pi\tilde{\varepsilon}_{\rm min}^{1/2}}\frac{|\varepsilon-\overline{U}|}{\tilde{\varepsilon}_{\rm min}}
\exp\left\{-\frac43\left|\frac{\varepsilon-\overline{U}}{\tilde{\varepsilon}_{\rm min}}\right|^{3/2}\right\},\\ \varepsilon-\overline{U}<0,\quad
1\ll\frac{|\varepsilon-\overline{U}|}{\tilde{\varepsilon}_{\rm min}}<|q_{\rm bi}|,
 \label{slo3}
\end{align}
\item Left plateau:
\begin{align}
 \rho(\varepsilon)\approx\rho_0.
 \label{slo4}
\end{align}
\end{enumerate}
The relevant energy scales $\overline{U}$ (shift of the peak) and $\tilde{\varepsilon}_{\rm min}$ (its width) are given by \eqref{bobo1a}, \eqref{dihy1fg} and \eqref{dihy1fgt}. For the tube and the strip cases these scales differ only in numerical prefactor. Logarithmically large parameter $q_{\rm bi}$ is defined in \eqref{dihy099}.

\section{Single impurity Non-Born effects: general results\label{non-Born general}}

In this section we turn to the discussion of single-impurity non-Born effects in resistivity of quasi-one-dimensional systems. The importance of non-Born effects in the systems with a singularity in the density of states was first discovered already in 60's in \cite{Fetter1965,MachidaShibata1972,Shiba1965,SodaMatsuuraNagaoka1967,Shiba1968} in the context of superconductors. In our recent paper \cite{IosPeshPRB2019} we have studied the very same problem of the non-Born effects in resistivity of conducting tube. We have shown that for $n\ll n_c$ one should account for non-Born renormalization of scattering amplitude $\Lambda^{\rm (ren)}$:
\begin{align}
\Lambda^{\rm (ren)}=\frac{\Lambda}{1+\Lambda \tilde{g}(\varepsilon)}.
\label{old_lambda}
\end{align}
Here $\tilde{g}(\varepsilon)$ is the difference between 2D and quasi-1D Green functions of ideal system. This formula holds down to $|\varepsilon|\sim n^2$, where the  scattering amplitude \eqref{old_lambda} is strongly suppressed, so that the multi-impurity effects become dominant and single-impurity approximation fails.

The result \eqref{old_lambda} was obtained in \cite{IosPeshPRB2019} by a solution of Dyson equation for scattering amplitude. Here we will introduce a more robust method that allows for description of non-Born scattering in general quasi-one-dimensional systems. Moreover, this approach can be conveniently generalized to take into account the essentially quantum multi-impurity effects and, therefore, to obtain the behavior of $\rho(\varepsilon)$ in the domain $|\varepsilon|\lesssim n^2$. The latter generalization, however, will be described in a separate publication. In this article we concentrate at the special effects that are absent in the case of a tube and exist in the case of a strip. 
These effects arise already in the energy range $|\varepsilon|\gg n^2$, where the single-impurity effects are still dominant and the semiclassical approach is sufficient.

For the case of point-like impurities the general dependence of matrix elements on $m$ and $k$ is as follows:
\begin{align}
V^{(i)}_{m_1k_1,m_2k_2}\equiv V^{(i)}_{m_1,m_2}e^{iz_i(k_1-k_2)},\\ V^{(i)}_{m_1,m_2}=\frac{\lambda}{\pi^2}\chi_{m_1}( x_i)\chi_{m_2}^{*}( x_i).
\label{nonres_scat_tubexx}
\end{align}

In order to evaluate the scattering rates for the current-carrying nonresonant states $m\neq N$,
 we may consider the corresponding self-energies $\Sigma_{mk}(\varepsilon)$:
\begin{align}
\tau^{-1}_{mk}=-2{\rm Im}\;\left\{\Sigma_{mk}\right\}.
\label{selfe1}
\end{align}

Within the Drude approximation the self-energy is additive with respect to different impurities; also it  depends on $k$ only through the total energy:
\begin{align}
\Sigma_{mk}=\sum_i\Sigma_{mk}^{(i)},\quad \Sigma_{mk}^{(i)}\equiv\Sigma_{m}^{(i)}\left(E=\varepsilon_m+\frac{k^2}{2m^*}\right)
\label{nonres_scat_tubexx0}
\end{align}
The self-energy can be expressed in terms of diagonal matrix elements of  renormalized scattering operator:
  \begin{align}
\Sigma^{(i)}_{m}=\tilde{V}^{(i){\rm (ren)}}_{m,m}.
\label{nonres_scat_tube1m}
\end{align}
Thus, our next task should be to evaluate $\tilde{V}^{(i){\rm (ren)}}_{m,m}$. 

To perform this evaluation we should single out the transitions involving states within the resonant band and take them into account nonperturbatively, while the transitions between nonresonant states can be treated perturbatively. For this purpose it is convenient to introduce a composite perturbative amplitude for transition between two nonresonant states $|m_1,k_1\rangle$, and $|m_2,k_2\rangle$ due to scattering at an impurity $i$:
\begin{align}
\tilde{V}^{(i)}_{m_1,m_2}=V^{(i)}_{m_1,m_2}+V^{(i)}_{m_1,N}G^{\rm (res)}_{\varepsilon}(z_i,z_i)V^{(i)} _{N,m_2}=\nonumber\\=\frac{\tilde{\lambda}_i}{\pi^2}\chi_{m_1}( x_i)\chi_{m_2}^{*}( x_i),\label{nonres_scat_tubepoll}\\
\tilde{\lambda}_i=\lambda\left\{1+\frac{\lambda_i}{\pi^2 }G^{\rm (res)}_{\varepsilon}(z_i,z_i)\right\},\label{nonres_scat_tubepoll1}\\ \lambda_i\equiv \lambda|\chi_{N}(x_i)|^2. 
\label{nonres_scat_tube}
\end{align}
  The first term in the right hand side of \eqref{nonres_scat_tubepoll1} describes the direct transitions between two nonresonant states, while the second term describes composite scattering processes with excursions to the resonant subband. Scattering processes that occur during the latter excursions are treated nonperturbatively in terms of the exact  Green function $G^{\rm (res)}_{\varepsilon}(z,z')$ for the purely one dimensional motion of an electron in the field of a single impurity.
 
   Now, to take into account multiple scattering processes, we should consider the following series for the renormalized matrix elements:
\begin{align}
\tilde{V}_{m_1,m_2}^{(i){\rm (ren)}}=\tilde{V}^{(i)}_{m_1,m_2}+\sum_{m\neq N}\tilde{V}^{(i)}_{m_1,m}g^{(m)}_{\varepsilon}(0)\tilde{V}^{(i)}_{m,m_2}+\nonumber\\+\sum_{m, m^{\prime}\neq N}\tilde{V}^{(i)}_{m_1,m}g^{(m)}_{\varepsilon}(0)\tilde{V}^{(i)}_{m,m^{\prime}}g^{(m^{\prime})}_{\varepsilon}(0)\tilde{V}^{(i)}_{m^{\prime},m_2}+\ldots ,\label{nonres_scat_tube1}
\end{align}
where
\begin{align}
g_{\varepsilon}^{(m)}(0)=\int\frac{dk}{2\pi}\left\{\varepsilon_m-\frac{k^2}{(2\pi)^2}+i0\right\}^{-1}=-\frac{\pi i}{\sqrt{\varepsilon_m}},
\end{align}
is the free one-dimensional Green function in the $m$th subband. The summation in \eqref{nonres_scat_tube1} runs over $m,m^{\prime}\ne N$ because all excursions to the resonant subband are already taken into account by the second term in \eqref{nonres_scat_tubepoll1}.

The series \eqref{nonres_scat_tube1} can be summed with the help of the Dyson equation
\begin{align}
\tilde{V}_{m_1,m_2}^{(i){\rm (ren)}}=\tilde{V}^{(i)}_{m_1,m_2}+\sum_{m\neq N}\tilde{V}^{(i)}_{m_1,m}g_{\varepsilon}^{(m)}(0)\tilde{V}^{(i){\rm (ren)}}_{m,m_2}.
\label{nonres_scat_tube2}
\end{align}
It is convenient to introduce a renormalized coupling constant $\tilde{\Lambda}^{{\rm (ren)}}_i$ according to
\begin{align}
\tilde{V}_{m_1,m_2}^{(i){\rm (ren)}}=\frac{\tilde{\Lambda}^{{\rm (ren)}}_i}{\pi^2}\chi_{m_1}( x_i)\chi_{m_2}^{*}( x_i),
\label{nonres_scat_tube2ed}
\end{align}
so that the scattering rate can be directly expressed through it:
\begin{align}
\tau^{-1}_{mk}=-\frac{2}{\pi^2}\sum_i|\chi_{m}( x_i)|^2{\rm Im}\;\left\{\tilde{\Lambda}^{{\rm (ren)}}_i\right\}=\nonumber\\=-\frac{2n}{\pi^2}\left\langle|\chi_{m}( x_i)|^2{\rm Im}\;\left\{\tilde{\Lambda}^{{\rm (ren)}}_i\right\}\right\rangle_{ x_i}
=\nonumber\\=-\frac{2n}{\pi^2}\int_0^1d x|\chi_{m}( x)|^2{\rm Im}\;\left\{\tilde{\Lambda}^{{\rm (ren)}}( x,\varepsilon)\right\},
\label{selfe1rew}
\end{align}
where we have used that the coupling constant $\tilde{\Lambda}^{{\rm (ren)}}_i=\tilde{\Lambda}^{{\rm (ren)}}( x_i,\varepsilon)$ depends on $i$ only through the transverse coordinate of impurity $ x_i$.

In terms of $\tilde{\Lambda}^{{\rm (ren)}}_i$ we can rewrite the Dyson equation \eqref{nonres_scat_tube2} as
\begin{align}
\tilde{\Lambda}^{{\rm (ren)}}_i=\tilde{\lambda}_i\left\{1+g_{\varepsilon}({\bf r}_i,{\bf r}_i)\tilde{\Lambda}^{{\rm (ren)}}_i\right\},\\
g_{\varepsilon}({\bf r}_i,{\bf r}_i)=\sum_{m\neq N}g_{\varepsilon}^{(m)}(0)|\chi_{m}( x_i)|^2,
\label{nonres_scat_tube3}
\end{align}
so that its solution is
\begin{align}
\tilde{\Lambda}^{{\rm (ren)}}_i=\frac{\tilde{\lambda}_i}{1-g_{\varepsilon}({\bf r}_i,{\bf r}_i)\tilde{\lambda}_i}.
\label{nonres_scat_tube4}
\end{align}
Having in mind that the resonant term $m=N$ is excluded from the summation in \eqref{nonres_scat_tube3}, in the semiclassical approximation ($N\gg 1$) we can relate $g_{\varepsilon}({\bf r},{\bf r}')$ to the Green function of a free two-dimensional electron and obtain
\begin{align}
g_{\varepsilon}({\bf r}_i,{\bf r}_i)\approx -i,\quad \tilde{\Lambda}^{{\rm (ren)}}_i=\frac{\tilde{\lambda}_i}{1+i\tilde{\lambda}_i}.
\label{nonres_scat_tube5}
\end{align}
Now, substituting \eqref{nonres_scat_tubepoll1} to \eqref{nonres_scat_tube5} and using $\lambda\ll 1$, we finally obtain  the renormalized scattering amplitude
\begin{align}
\tilde{\Lambda}^{{\rm (ren)}}_i=\Lambda\left\{1+\frac{\Lambda_i}{\pi^2Q_i+\Lambda_i^*}\right\},\label{nonres_scat_tube5as}\\ Q_i=\left[G^{\rm (res)}_{\varepsilon}(z_i,z_i)\right]^{-1}-\lambda_i/\pi^2,
\label{nonres_scat_tube5ask}
\end{align}
where
\begin{align}
\Lambda=\frac{\lambda}{1+i\lambda}\approx\lambda-i\lambda^2, \quad \Lambda_i=\Lambda|\chi_{N}( x_i)|^2,
\label{nonres_scat_tube5as1}
\end{align}
$\Lambda$ being the complex scattering amplitude for an electron in an infinite plane. Note that it satisfies the two-dimensional version of the optical theorem (see \cite{opticaltheorem}):
\begin{align}
{\rm Im}\;\Lambda=-|\Lambda|^2.
\label{nonres_scat_tube5as1}
\end{align}

We have to  stress again that the result \eqref{nonres_scat_tube5as} is only valid within the single-impurity approximation which  is correct in the  semiclassical range of energies $n^2\ll\varepsilon$.

Now we proceed with evaluating $\tilde{\lambda}_i$. 
Generally speaking, $G^{\rm (res)}_{\varepsilon}(z,z_i)$ satisfies the following equation:
\begin{align}
\left\{-\frac{d^2}{(2\pi)^2dz^2}+\sum_j\frac{\lambda_j}{\pi^2}\delta(z-z_j)-\varepsilon\right\}G_{\varepsilon}^{\rm (res)}(z,z_i)=\nonumber\\=-\delta(z-z_i),
\label{scat_ampl_tube_general}
\end{align}
where the summation runs over all impurities.
However, in this paper we restrict our consideration by the single-impurity approximation, which allows to consider only one impurity and discard all the terms in the sum except one with $j=i$. This approximation is only justified under the semiclassical condition, when the typical wave-lengths $\lambda_F^{(1D)}\sim \varepsilon^{-1/2}$ of one-dimensional electron wave-functions within the resonant band are much shorter than the typical distance $\Delta z\sim n^{-1}$ between impurities. So, the semiclassical condition $\lambda_F^{(1D)}\ll\Delta z$ can be written as
\begin{align}
|\varepsilon|\gg \varepsilon_{\min}^{\rm (nB)}\sim n^2.
\label{scat_ampl_semic}
\end{align}
In Section \ref{Decay rates for resonant states and Breakdown of single-impurity approximation} we will show that the same condition  also arises from the requirement $\tau_{\rm res}^{-1}\ll\varepsilon$, which means that the corresponding states are well-defined.
Thus, in the energy range \eqref{scat_ampl_semic} the equation \eqref{scat_ampl_tube_general} may be rewritten as
\begin{align}
\left\{-\frac{d^2}{(2\pi)^2dz^2}+\frac{\lambda_i}{\pi^2}\delta(z)-\varepsilon\right\}G_{\varepsilon}^{\rm (res)}(z,0)=-\delta(z),
\label{scat_ampl_tube00}
\end{align}
where $z_i$ was chosen at the origin ($z_i=0$). Note that the minus sign appeared at the right hand side due to the standard definition $\hat{G}=(\varepsilon - \hat{H})^{-1}$. The solution of \eqref{scat_ampl_tube00} leads to
\begin{align}
G_{\varepsilon}(z,0)=\frac{\pi e^{2\pi i\sqrt{\varepsilon}|z|}}
{i\sqrt{\varepsilon}-\lambda_i/\pi},\quad G_{\varepsilon}(0,0)=\frac{\pi}{i\sqrt{\varepsilon}-\lambda_i/\pi},\label{scat_ampl_tube10edr}\\ \tilde{\lambda}_i=\frac{i\lambda\sqrt{\varepsilon}}
{i\sqrt{\varepsilon}-\lambda_i/\pi},\quad Q_i=i\sqrt{\varepsilon}/\pi-2\lambda_i/\pi^2.
\label{scat_ampl_tube10}
\end{align}
Now, using \eqref{scat_ampl_tube10} we can rewrite the result \eqref{nonres_scat_tube5as} in the form
\begin{align}
\tilde{\Lambda}^{{\rm (ren)}}_i=\tilde{\Lambda}^{{\rm (ren)}}(\epsilon,t_i)=\frac{\sqrt{-\epsilon}|\lambda|(1-i\lambda)}{\sqrt{-\epsilon}\,{\rm sign}\lambda-(1-i\lambda)2t_i},
\label{nonres_scat_tube5as2}
\end{align}
where $t_i$ is defined by \eqref{nonres_scat_tube6ewre} and we have introduced
\begin{align}
\epsilon=\varepsilon/\varepsilon_{\rm nB}\label{defi}
\end{align}
for brevity.

We see that the single impurity non-Born scattering effects are most spectacular for $\epsilon\ll1$ since for $\epsilon\to 0$ $\tilde{\Lambda}^{\rm (ren)}\to 0$. Thus, if one compares $\varepsilon_{\rm nB}$ with $\varepsilon_{\rm min}$ from the Section\ref{Smoothing of the singularity: qualitative description}, the following criterion for the single-impurity non-Born effects to come into play earlier than the multi-impurity ones can be obtained:
\begin{align}
\varepsilon^{\rm (B)}_{\rm min}<\varepsilon_{\rm nB},\quad\mbox{or}\quad n<n_c.
\end{align}
In what follows we will concentrate on the non-Born case $n<n_c$. 

Now, taking imaginary part of both sides of \eqref{nonres_scat_tube5as2} and expanding it up to the second order in $\lambda$, we obtain:
\begin{align}
-{\rm Im}\;\{\tilde{\Lambda}^{{\rm (ren)}}_i\}\approx \left\{\begin{aligned}\frac{2t_i\sqrt{\epsilon}|\lambda|}{\epsilon+4t_i^2}-\lambda^2\frac{\epsilon(4t_i^2-\epsilon)}{(4t_i^2+\epsilon)^2},& \quad(\varepsilon>0),\\
\frac{-\lambda^2\epsilon}{\left(2t_i+{\rm sign}\,\lambda\sqrt{-\epsilon}\right)^2+4t_i^2\lambda^2},& \quad(\varepsilon<0).
\end{aligned}\right.
\label{nonres_scat_tube6}
\end{align}
Formulas \eqref{nonres_scat_tube6} and \eqref{selfe1rew} determine the scattering rate for any current-carrying state, characterized by the energy $\varepsilon$ and the subband index $m$.

For $|\epsilon_i|\gg1$ both upper and lower lines in \eqref{nonres_scat_tube6}, as expected, are reduced to the trivial 2D result: ${\rm Im}\;\{\tilde{\Lambda}_i^{{\rm (ren)}}\}=-\lambda^2$. However, the second term in the upper line of \eqref{nonres_scat_tube6} (proportional to $\lambda^2$) starts to dominate over the first one only at very large $\epsilon_i\gtrsim\lambda^{-2}\gg 1$, i.e.,  away from the Van Hove singularity, where the contribution of the resonant subband is already irrelevant.
In what follows we will mainly stick to the most interesting case $\epsilon_i\lesssim1$ when one can discard the second term  in the upper line of \eqref{nonres_scat_tube6} and write: 
\begin{align}
-{\rm Im}\;\{\tilde{\Lambda}^{{\rm (ren)}}(\epsilon,t)\}\approx\nonumber\\\approx \left\{\begin{aligned}\frac{2t_i\sqrt{\epsilon}|\lambda|}{\epsilon+4t^4},& \quad(\varepsilon>0),\\
\frac{-\lambda^2\epsilon}{\left(2t+{\rm sign}\,\lambda\sqrt{-\epsilon}\right)^2+4t^2\lambda^2},& \quad(\varepsilon<0).
\end{aligned}\right.
\label{nonres_scat_tube66}
\end{align}

Similar to the case of tube (see \cite{IosPeshPRB2019}), for $\lambda<0$ the expression \eqref{nonres_scat_tube6} has a sharp maximum at $\varepsilon=\varepsilon_i^{\rm qs}$
where
\begin{align}
\varepsilon_i^{\rm qs}=-(\lambda_i/\pi)^2=-4t_i^2\varepsilon_{\rm nB}.
\label{nonres_scat_tube7bwe}
\end{align}
Actually, a formal expansion of \eqref{nonres_scat_tube5as2} up to the second order in $\lambda$ leads to a result divergent at $\varepsilon\to\varepsilon_i^{\rm qs}$. However, more accurate calculations give rise to an additional term $4t_i^2\lambda^2$ in the denominator of lower-line formula \eqref{nonres_scat_tube66}. This modification regularizes the divergency. 

As we have shown in \cite{IosPeshPRB2019}, this $\varepsilon_i^{\rm qs}$ is nothing else but the energy of a quasistationary state that is formed near any attractive impurity in any quasi-one-dimensional system. What is essentially new in the strip case, compared to the tube one -- here the energy $\varepsilon_i^{\rm qs}$ is not unique, but depends on the position  of impurity $ x_i$. This dependence, as we will see soon, leads to inhomogeneous broadening of the resonant peak.

\section{Non-Born scattering rate and resistivity: general results\label{Non-Born scattering rate and resistivity: general results}}

Now we are prepared to write down explicit expressions for scattering rates of current-carrying nonresonant states $\tau^{-1}_{m}$ and the resistivity $\rho$ with the help of \eqref{nonres_scat_tube66},\eqref{selfe1rew}:
\begin{align}
\frac{\tau_0}{\tau_m(\varepsilon)}=-\frac{1}{\lambda^2}{\rm Im}\;\int_0^1 d x|\chi_m( x)|^2{\rm Im}\;\{\tilde{\Lambda}^{\rm (ren)}( x,\varepsilon)\}=\nonumber\\=-\frac{1}{\lambda^2}\int_0^1 d x(1-\cos (2\pi m x)){\rm Im}\;\{\tilde{\Lambda}^{\rm (ren)}( x,\varepsilon)\}.
\label{res_repuls0}
\end{align}
It is easy to understand that the second, $m$-dependent term in \eqref{res_repuls0} vanishes after integration over $ x$. Indeed, $\tilde{\Lambda}^{\rm (ren)}( x,\varepsilon)$ depends on $ x$ only in the form of combination $\cos 2\pi N x$. It means that the integrand of \eqref{res_repuls0} can be written as a Fourier series
\begin{align}
\sum_{l=0}^\infty(1-\cos (2\pi m x))A_l\cos (2\pi lN x)
\label{res_repuls01}
\end{align}
with certain coefficients $A_l$. The only term in this series that survives the integration over $ x$ is $A_0$, and it is $m$-independent! Thus, we conclude that $\tau_m\equiv\tau_{\rm nonres}(\varepsilon)$ does not depend on $m$, and it is true not only within the Born approximation but also beyond. Then
\begin{align}
\frac{\rho(\varepsilon)}{\rho_0}=\frac{\tau_0}{\tau_{\rm nonres}(\varepsilon)}=-\frac{1}{\lambda^2}\int_0^1 d x{\rm Im}\;\{\tilde{\Lambda}^{\rm (ren)}( x,\varepsilon)\}.
\label{res_repuls02e}
\end{align}
Since the integrand of \eqref{res_repuls02e} is periodic function of $ x$ with period 1, it is more convenient to perform the averaging in terms of variables $t_i$ instead of $ x_i$. For the case of strip, using explicit expression
\begin{align}
t_i=\sin^2(\pi N x_i),\label{res_repuls03q}
\end{align}
we arrive at
\begin{align}
\frac{\rho(\varepsilon)}{\rho_0}=-\frac{1}{\pi \lambda^2}\int_0^1 \frac{dt}{\sqrt{t(1-t)}}{\rm Im}\;\{\tilde{\Lambda}^{\rm (ren)}(\varepsilon,t)\}
\label{res_repuls02}
\end{align}
and substituting \eqref{nonres_scat_tube66} 
 to \eqref{res_repuls02e} we obtain:
\begin{align}
\frac{\rho(\varepsilon)}{\rho_0}=
\begin{cases}\frac{1}{|\lambda|}F_1(\epsilon),\quad\epsilon>0,\\
F_2(\epsilon,\lambda),\quad \epsilon<0,
\end{cases}
\label{res_repuls03}
\end{align}
where
\begin{align}
F_1(\epsilon)
=\int_0^1\frac{\sqrt{\epsilon}}{\epsilon+4t^2}\frac{2tdt}{\pi\sqrt{t(1-t)}}=\left(\frac{\sqrt{1+4/\epsilon}-1}{2(1+4/\epsilon)}\right)^{1/2},
\label{res_repuls}
\end{align}
\begin{align}
F_2(\epsilon,\lambda)
=\int_0^1\frac{-\epsilon}{({\rm sign}\,\lambda \sqrt{-\epsilon}+2t)^2+4t^2\lambda^2}\frac{dt}{\pi\sqrt{t(1-t)}}.\label{rho_attract}
\end{align}
Functions $F_1(\epsilon)$,  $F_2(\epsilon,\lambda)$ are evaluated in Appendix. The second term in the denominator in \eqref{rho_attract} originates from the similar term in \eqref{nonres_scat_tube6}, it is only essential for attracting impurities ($\lambda<0$) and only in the vicinity of quasistationary resonance. For repulsing impurities this term can be altogether neglected; in this case a separate dependence of $F_2$ on $\lambda$ vanishes: $F_2(\epsilon,\lambda)\to F_2(\epsilon)$.

\section{Non-Born resistivity: repulsing impurities\label{Non-Born resistivity: repulsing impurities}}
It this section we will analyze the general results obtained above for the case of repulsive impurities, $\lambda>0$.  For $\epsilon>0$  we find (see Appendix \ref{Evaluation of function1})
\begin{align}
\frac{\rho(\epsilon)}{\rho_{0}}=\frac{1}{\lambda}\;
\left(\frac{\sqrt{1+4/\epsilon}-1}{2(1+4/\epsilon)}\right)^{1/2}\approx\nonumber\\\approx\frac{1}{\lambda}\;\left\{
	\begin{aligned}
	\epsilon^{1/4}/2,\quad	&\mbox{for $\epsilon\ll 1$},
\\1/\sqrt{\epsilon},\quad  & \mbox{for $\epsilon\gg 1$},
	\end{aligned}\right.
\label{rho_rep_energy+}		
\end{align}
As we have already mentioned in the previous section, for $\lambda>0$ the formula for $F_2$ is simplified and for $\epsilon<0$ we obtain (see Appendix \ref{Evaluation of function2r})
\begin{align}
\frac{\rho(\epsilon)}{\rho_{0}}\approx F_2(\epsilon, 0)=\nonumber\\=\frac{(-\epsilon)^{1/4}(1+(-\epsilon)^{1/2})}{(2+(-\epsilon)^{1/2})^{3/2}}\approx\left\{
	\begin{aligned}
\frac{1}{2\sqrt{2}}(-\epsilon)^{1/4},\quad  & \mbox{for $|\epsilon|\ll1$},\\
1,\quad	&\mbox{for $|\epsilon|\gg1$},
	\end{aligned}\right.
\label{rho_rep_energy-}		
\end{align}
The maximum
\begin{align}
\frac{\rho^{(+)}_{\max}}{\rho_{0}}=\frac{1}{2\sqrt{2}\lambda}
\label{self-consisyy}		
\end{align}
is reached at $\varepsilon=\frac43\varepsilon_{\rm nB}$. Thus, the maximum of the resistivity, observed at $\varepsilon>0$, in the case of strip is somewhat broadened, compared to that in the case of tube. The overall $\rho(\epsilon)$ dependence for repulsing impurities is shown in. Fig.\ref{nonBornpos} for both cases of a tube and a strip.

\subsection{Paradox:  weak impurities scatter more effectively than strong ones! \label{Paradox}}

It is  important to note a different (compared to the case of tube) law $\rho\propto|\epsilon|^{1/4}$ (\ref{rho_rep_energy+}), (\ref{rho_rep_energy-}) of vanishing $\rho(\epsilon)$ at $|\epsilon|\to 0$ for both signs of $\epsilon$. For the tube the analogous law is $\rho\propto|\epsilon|^{1/2}$. To elucidate the reason for this difference let us analyze the integral over $t$ in \eqref{res_repuls}. While for $|\epsilon|\gtrsim 1$ the entire interval $0< x<1$ (or $t\sim 1$) contributes to this integral, for $|\epsilon|\ll 1$ the main contribution comes from small
\begin{align}
t\sim \sqrt{|\epsilon|}\ll 1.
\label{self-conso0}		
\end{align}
It means that scattering at ``weak'' impurities, situated close to nodes of the transversal wave-function of the resonant band, turns out to be more effective than scattering at the strong ones, sitting close to antinodes. How it can possibly be? 

The physical reason is the following. For small $t$, characteristic for weak impurities, the scattering of slow particles with small $\epsilon\sim t^2$ is strongly enhanced due to the resonance at virtual level, lying  at $\epsilon=-4t^2$ on the unphysical sheet of the Riemann surface of complex $\epsilon$. As a result, for given  $\epsilon$ the most efficient scatterers are weak impurities with $t\sim\sqrt{\epsilon}/2$. Thus, the scattering at the impurities with large bare $\Lambda_i\gg\sqrt{|\epsilon|}$ turns out to be suppressed stronger than scattering at those with moderately small
\begin{align}
\Lambda_i\sim\sqrt{|\epsilon|}\ll 1.
\label{self-consooo}		
\end{align}

Thus, we arrive at paradoxical and exciting conclusion: though for small $|\epsilon|\ll 1$ the scattering is generally suppressed, the residual weak scattering is dominated by presumably ineffective impurities, that sit relatively close to the nodes, at distances $ x_i\sim\lambda_F|\epsilon|^{1/4}\ll \lambda_F$ ($\lambda_F\sim1/N$ being the Fermi wavelength, or the distance between the neighboring  nodes) and have, therefore, anomalously small bare scattering amplitudes. As one of the consequences, the resistivity of a strip vanishes with $|\epsilon|\to 0$ slower than the resistivity of a tube.

\section{Non-Born resistivity: attracting impurities\label{attracting impurities0}}

Above the Van Hove singularity, for $\varepsilon>0$ the scattering rate depends only on $\lambda^2$, so that the case of attracting impurities does not differ from that of the repulsing ones and the resistivity for $\varepsilon>0$ is described by the formula \eqref{rho_rep_energy+}. Below the Van Hove singularity, for $\varepsilon<0$, however, there are some impressive effects, specific for the attractive impurities. They are mostly due to the presence of quasistationary states.

\subsection{Quasistationary states: ``impurity band''.\label{Attracting impurities1}}

As we have shown in \cite{IosPeshJETPL2018,IosPeshPRB2019}, in  a quasi-one-dimensional system each attracting impurity forms a quasistationary state below each subband of transverse quantization. These states arise for arbitrary weak attraction, without a threshold. Moreover, for  weak attraction the quasistationary states are even better defined than for strong one: the quality factor (i.e., the ratio of the energy  to the decay rate) increases with decreasing strength of attraction. The quasistationary states are manifested as poles of the renormalized scattering amplitude
in the complex $\epsilon$ plane.

 In contrast to the case of cylinder,  to each impurity $i$ in a strip corresponds its own value of the scattering amplitude $\Lambda_i\approx (\lambda-i\lambda^2)2t_i^2$, so that energies of the quasistationary states are different at different impurities:
\begin{align}
\epsilon_{\rm qs}(t)=4t^2(-1+2i\lambda),\quad \varepsilon_{\rm qs}=\epsilon_{\rm qs}\varepsilon_{\rm{nB}}.
\label{self-conser33}		
\end{align}
Let's forget for a while about small imaginary part of  $\epsilon_{\rm qs}$; we will easily restore it in a due time.
We see that values of $\epsilon_{\rm qs}$ are confined in a sort of ``impurity band'' that spans an interval of energies $-4<\epsilon_{\rm qs}<0$. Since $ x_i$ is a random variable homogeneously distributed between 0 and 1, the ``density of quasistationary states'', i.e., the distribution function for $\varepsilon_{\rm qs}$, is
 \begin{align}
P(\epsilon_{\rm qs})=\int_0^1\frac{dt}{\pi\sqrt{t(1-t)}}\delta\left[\epsilon_{\rm qs}+4t^2\right]=\nonumber\\=\frac{\theta(-\epsilon_{\rm qs})\theta(\epsilon_{\rm qs}+4)}{2\pi}\frac{\left(2+\sqrt{-\epsilon_{\rm qs}}\right)^{1/2}}{\sqrt{(-\epsilon_{\rm qs})^{3/2}(4+\epsilon_{\rm qs})}}.\label{dedise}
\end{align}
Thus, outside the impurity band, for $\epsilon<-4$ the scattering is only possible to usual states of continuous spectrum, while within the impurity band, for $-4<\epsilon<0$, in principle, both continuum and quasistationary states may serve as final states of scattering processes. In fact, we will see that quasistationary states dominate everywhere in this range, except narrow interval at the edge of the impurity band, at $\epsilon=-4$, with a width 
 \begin{align}
\Gamma\sim{\rm Im}\,\epsilon_{\rm qs}\sim |\lambda|
\label{dedisea}
\end{align}
being of order of the decay rate for the quasistationary states.

\subsection{Scattering outside the impurity band.\label{Far tail}}

Since there are no quasistationary states in the energy range $\epsilon<-4$, here we can simply put $\lambda=0$ in \eqref{rho_attract}. The corresponding integral is evaluated in Appendix \ref{Evaluation of function2a} (see \eqref{deni2pp4s100i}) and we get
\begin{align}
\frac{\rho(\epsilon)}{\rho_{0}}\approx F_2(\epsilon, 0)=\frac{(-\epsilon)^{1/4}\left(\sqrt{-\epsilon}-1\right)}{\left(\sqrt{-\epsilon}-2\right)^{3/2}}\approx\nonumber\\\approx\left\{\begin{aligned}8\sqrt{2}\left[-(\epsilon+4)\right]^{-3/2}, & \quad\mbox{for $-(\epsilon+4)\ll 1$,}\\
1,& \quad \mbox{for $|\epsilon|\gg 1$.}
\end{aligned}
\right.
 \label{deni2pp4s99}
\end{align}

\subsection{Scattering within the impurity band.\label{Near tail}}

For any given energy in the range $-4<\epsilon<0$ the leading contribution to the resistivity comes from the scattering on resonant impurities with such $t_i$ that $\epsilon_{\rm qs}(t_i)\approx \epsilon$. In contrast with the previous case, to avoid divergency, here we have to take into account the imaginary part of $\epsilon_{\rm qs}(t)$. The corresponding calculations are presented in Appendix \ref{Evaluation of function2a1}, resulting in \eqref{leeft1}:
\begin{align}
\frac{\rho(\epsilon)}{\rho_{0}}=F_2(\epsilon, \lambda)=\frac{1}{|\lambda|}\left(\frac{\sqrt{-\epsilon}}{2-\sqrt{-\epsilon}}\right)^{1/2}\approx\nonumber\\\approx\frac{1}{|\lambda|\sqrt{2}}\left\{
	\begin{aligned}
(-\epsilon)^{1/4},\quad  & \mbox{for $|\epsilon|\ll 1$},\\
4(4+\epsilon)^{-1/2},\quad	&\mbox{for $4+\epsilon\ll 1$}.
	\end{aligned}\right.\label{deni2pp4po33ws}
\end{align}
\subsection{Van Hove-like feature at the edge of impurity band: scattering at strongest impurities.\label{Sharp peak}}

Combining \eqref{deni2pp4s99} and \eqref{deni2pp4po33ws}, we arrive at
 \begin{align}
\frac{\rho(\epsilon)}{\rho_{0}}=\left\{
	\begin{aligned}
8\sqrt{2}\left[-(\epsilon+4)\right]^{-3/2},\quad  & \mbox{for $4+\epsilon\to -0$},\\
\frac{2\sqrt{2}}{|\lambda|}(4+\epsilon)^{-1/2},\quad	&\mbox{for $4+\epsilon\to +0$}.
	\end{aligned}\right.\label{deni2pp4po33wsee}
\end{align}
 Thus, at the lower edge of impurity band, at $\epsilon=-4$ the resistivity has an asymmetric (formally divergent) peak, somewhat similar to Van Hove singularity.

 This entire feature is nothing else, but the inhomogeneously broadened (due to the dispersion of scattering amplitudes $\lambda_i$ for different impurities) peak of the resonant scattering, which in the case of cylinder (where all $\lambda_i$  are identical) was manifested as a sharp line (see \cite{IosPeshJETPL2018,IosPeshPRB2019}).

 The Van Hove-like singularity in resistivity \eqref{deni2pp4po33wsee} reflects  just the divergency of the density of quasistationary states $P(\epsilon)$ at the edge of the impurity band. Since $\min\epsilon_{\rm qs}=-4$ is reached for $t=1$, we see that scattering near the peak is dominated by strongest impurities, sitting near the antinodes of transversal wave-function.

 This Van Hove-like singularity at $|\epsilon|\to 4$ is indeed smeared in the range $|\epsilon+4|\lesssim|\lambda|$, where the contributions of both types of final states --  the continuum and the quasistationary states -- are comparable. To elucidate this mixing one should treat the integral in Eq\eqref{rho_attract} more accurately, without using an approximate formula \eqref{deni2pp4po3pp}.  As a result of calculations (see Appendix \ref{Evaluation of function2a2}) we obtain
\begin{align}
\frac{\rho(\epsilon)}{\rho_{0}}=F_2(\epsilon,\lambda)=\frac{1}{\sqrt{2}}\left(\frac{\sqrt{a^2+1}-a}{|\lambda|^3(a^2+1)}\right)^{1/2}\approx\nonumber\\\approx\left\{\begin{aligned}\frac{8\sqrt{2}}{[-(\epsilon+4)]^{3/2}},&\quad \mbox{for $8|\lambda|\ll -(\epsilon+4) \ll1$},\\
 \frac{2\sqrt{2}}{|\lambda|(4+\epsilon)^{1/2}},&\quad \mbox{for $8|\lambda|\ll 4+\epsilon \ll1$},
 \end{aligned}\right.\label{deni2p76}
\end{align}
where $a=-(\epsilon+4)/8|\lambda|$. Naturally, the asymptotics of \eqref{deni2pp4po33wsee} and \eqref{deni2p76} overlap at $|\lambda|\ll|\epsilon+4|\ll 1$. The function $F_2$ reaches its maximum $F_2^{(\max)}=\frac{3^{3/4}}{2\sqrt{2}|\lambda|^{3/2}}$ at $a=-3^{-1/2}$, so that the maximal resistivity
\begin{align}
\frac{\rho^{(-)}_{\max}}{\rho_{0}}=\frac{3^{3/4}}{2\sqrt{2}|\lambda|^{3/2}}
\label{self-consisyyi}		
\end{align}
is reached at $\epsilon=-4\left(1-\frac{2|\lambda|}{\sqrt{3}}\right)$. The width of this maximum $\Gamma\sim |\lambda|\ll 1$. 

Thus, we conclude that the left peak of resistivity (that exists only for attracting impurities) is higher than the right one: its height is proportional to $|\lambda|^{-3/2}$ instead of $|\lambda|^{-1}$. On the other hand, due to the inhomogeneous broadening,    it is lower than it would be in the case of cylinder: $|\lambda|^{-3/2}$ instead of $|\lambda|^{-2}$.

\subsection{Low energy scattering at weak impurities.\label{Low energy }}

In contrast with the case of lower edge of the impurity band (at $\epsilon\to -4$), the divergency of $P(\epsilon)$ at the upper edge (i.e., at $\epsilon\to 0$) does not lead to divergency of  $\rho(\epsilon)$. The divergency of $P(\epsilon)$ appears to be not strong enough to overcome the tendency for ${\rm Im}\,\Lambda^{\rm (ren)}$ to vanish due to non-Born screening. As a result, for attracting impurities
$\rho(\epsilon)$ still goes to zero at $\epsilon\to-0$, but for all energies $\rho(\epsilon)$ is much larger than that in the case of repulsing impurities:  $\rho_{\rm attr}(\epsilon)\gg \rho_{\rm rep}(\epsilon)$. The strong resonant scattering at quasistationary states with low binding energies $\epsilon_{\rm qs}\approx \epsilon$ gives additional large factor $|\lambda|^{-1}$ in $\rho(\epsilon)$ dependence at $\epsilon<0$, $|\epsilon|\ll 1$:  
\begin{align}
\frac{\rho(\epsilon)}{\rho_{0}}=\frac{\tau_0}{\tau_{\rm nonres}(\epsilon)}\approx\frac{|\epsilon|^{1/4}}{\sqrt{2}}\left\{\begin{aligned}
1/|\lambda|,&\quad\mbox{for  $\lambda<0$, }\\
1/2,&\quad\mbox{for $\lambda>0$.}
\end{aligned}\right.
\label{deni2pp4po33ws2}
\end{align}
Here we stress again that the scattering at $|\epsilon|\ll 1$ (for both $\epsilon>0$ and $\epsilon<0$) is dominated by weak impurities, sitting close to nodes of the transversal wave functions.

\section{Decay rates for resonant states and Breakdown of single-impurity approximation\label{Decay rates for resonant states and Breakdown of single-impurity approximation}}

What is the boundary energy $\varepsilon_{\rm min}^{\rm(nB)}$ below which the above theory breaks down due to effects of multi-impurity scattering? 

The current carrying nonresonant states themselves have large kinetic energy ($\varepsilon_m\gtrsim N$ in our units) so that the semiclassical condition for them $\tau_{\rm nonres}^{-1}\ll \varepsilon_m$ is granted. However, for correct evaluation of the resistivity $\rho(\varepsilon)$ we need reliable expressions for the scattering rates $\tau_{\rm nonres}^{-1}(\varepsilon)$ of these states. As we have seen in previous sections,  intermediate resonant states with kinetic energies $\varepsilon'\sim |\varepsilon|$ play crucial role in these scattering processes. Thus, we should require that not only the nonresonant states, but also resonant ones with relevant energies are well defined: $\tau_{\rm res}^{-1}(|\varepsilon|)\ll |\varepsilon|$.

So,  we have to evaluate the scattering rate $\tau_{\rm res}^{-1}(\varepsilon')$ for the states in the resonant band with kinetic energy $\varepsilon'\sim |\varepsilon|>0$. Therefore $\varepsilon_{\rm min}^{\rm(nB)}$ should be found from the estimate
\begin{align}
\tau_{\rm res}^{-1}(\varepsilon_{\rm min}^{\rm(nB)})\sim\varepsilon_{\rm min}^{\rm(nB)}.
\label{res_ampl_pert0cv}
\end{align}
The same criterion we have already used in the Born case, when $n\gg n_c$ (see Section section \ref{Smoothing of the singularity: qualitative description}), and it has led us to the result $\varepsilon_{\rm min}^{\rm(B)}\sim(n\lambda^2)^{2/3}$ there. However, in the Born case $\tau_{\rm res}^{-1}(\varepsilon')\approx (3/2) \tau_{\rm nonres}^{-1}(\varepsilon')$ and we did not have to make a separate calculation for
$\tau_{\rm res}^{-1}(\varepsilon')$. In the present non-Born regime, as we will see,  $\tau_{\rm res}^{-1}(\varepsilon')\ll \tau_{\rm nonres}^{-1}(\varepsilon')$ for $\varepsilon'\ll\varepsilon_{\rm nB}$ and such a separate calculation is necessary. For brevity in the rest of this section we will  write simply 
$\varepsilon$ instead of $|\varepsilon|$ having in mind that thus defined $\varepsilon$ is necessarily positive.

Evaluation of scattering rate for states within the resonant band technically may be performed in a way very similar to that which we have used in Section \ref{Non-Born scattering rate and resistivity: general results} for current-carrying nonresonant states. Both rates are governed by the same renormalized scattering amplitude $\tilde{\Lambda}^{\rm (ren)}( x,\varepsilon)$, the only difference is the change of the prefactor $|\chi_m( x)|^2\to|\chi_N( x)|^2$ in formula \eqref{selfe1rew}. This modification, however, turns out to have very serious consequences for $\epsilon\ll 1$. 
\begin{align}
\frac{1}{\tau_{\rm res}(\varepsilon)}\equiv\frac{1}{\tau_{N}(\varepsilon)}=\nonumber\\=-2\sum_i{\rm Im}\,\left\{\Sigma^{(i)}_N(\varepsilon)\right\}=-2\sum_i{\rm Im}\,\left\{\tilde{V}_{NN}^{(i){\rm (ren)}}\right\},
\label{res_ampl_pert0}
\end{align}
\begin{align}
\frac{\tau_0}{\tau_{\rm res}(\varepsilon)}=-\frac{1}{\lambda^2}\int_0^1 d x|\chi_N( x)|^2{\rm Im}\;\{\tilde{\Lambda}^{\rm (ren)}( x,\varepsilon)\}=\nonumber\\=
-\frac{1}{\pi \lambda^2}\int_0^1 \frac{2tdt}{\sqrt{t(1-t)}}{\rm Im}\;\{\tilde{\Lambda}^{\rm (ren)}(\varepsilon,t)\}=
\frac{\tilde{F}_1(\epsilon)}{|\lambda|},
\label{res_repuls02}
\end{align}
where
\begin{align}
\tilde{F}_1(\epsilon)
=\int_0^1\frac{\sqrt{\epsilon}}{\epsilon+4t^2}\frac{(2t)^2dt}{\pi\sqrt{t(1-t)}}=\nonumber\\=\sqrt{\epsilon}\left(1-\sqrt{\frac{1+\sqrt{1+4/\epsilon}}{2(1+4/\epsilon)}}\right).
\label{res_repulsr}
\end{align}
The expression for $\tilde{F}_1(\epsilon)$ differs from expression \eqref{res_repuls} for $F_1(\epsilon)$ by an extra factor $2t$ in the integrand.  Evaluation of function $\tilde{F}_1(\epsilon)$ for general $\epsilon>0$ is performed in Appendix \ref{Evaluation of function1}. We actually need only its behaviour at $\epsilon\ll 1$:
 \begin{align}
\tilde{F}_1(\epsilon)\approx\sqrt{\epsilon},
\label{Fneg}
\end{align}
so that, for $\varepsilon\ll\varepsilon_{\rm nB}$ we obtain
\begin{align}
\frac{1}{\tau_{\rm res}(\varepsilon)}=\frac{1}{\tau_0}\frac{\pi\sqrt{\varepsilon}}{\lambda^2}=\frac{2n}{\pi}\sqrt{\varepsilon}\ll \frac{1}{\tau_{\rm nonres}(\varepsilon)}.
\label{res_repuls03r}
\end{align}
This drastic difference in behaviour of $\tau_{\rm res}^{-1}(\varepsilon)$ and $\tau_{\rm nonres}^{-1}(\varepsilon)$ at $\epsilon\ll 1$ can be explained in the following way: In contrast with the integral \eqref{res_repuls} that converges at $t\sim\sqrt{\epsilon}\ll 1$, the integral \eqref{res_repulsr}, due to additional factor $2t$, converges at $t\sim 1$. Physically, it means that while the scattering of nonresonant states is dominated by weak impurities, for the scattering of resonant states  weak impurities do not play any distinguished role: all typical impurities contribute to scattering of resonant states equally. It also explains why the result \eqref{res_repuls03r} does not differ from similar result obtained in \cite{IosPeshPRB2019} for the case of tube, where all the impurities were equivalent.

Substitution of \eqref{res_repuls03r} to criterion \eqref{res_ampl_pert0cv} gives
$\varepsilon_{\rm min}^{\rm (nB)}\sim n^2$
which is also in accord with the case of tube. Now we can conclude that the results obtained in sections \ref{non-Born general}-\ref{attracting impurities0} are valid  in the energy range
\begin{align}
|\varepsilon|\gg\varepsilon_{\rm min}^{\rm (nB)}.
\label{res_repuls03oo}
\end{align}
In the range $|\varepsilon|\lesssim \varepsilon_{\rm min}^{\rm (nB)}$, however, the behavior of $\rho(\varepsilon)$ can be studied with the help of some generalized approach. It takes into account multi-impurity effects, though only for scattering within the resonant band  and is based upon using some exact results from the theory of strictly one-dimensional systems. These results will be discussed elsewhere.

\section{conclusion.\label{Discussion and conclusion}}

In our previous paper on the role of non-Born effects in resistivity of metallically conducting tubes \cite{IosPeshPRB2019} we have shown that there exists certain crossover concentration of impurities $n_c$ and studied the non-Born effects for both $n\gg n_c$ and $n\ll n_c$. In this paper we have modified our approach to refine the results of \cite{IosPeshPRB2019} and -- most important -- extended it to the case of  ``strips'' -- constrictions in two-dimensional systems.

For $n\gg n_c$ we were able to find the resistivity $\rho(\varepsilon)$ in the entire range of $\varepsilon$. For $|\varepsilon|\gg\varepsilon_{\rm min}^{\rm (B)}$ this is a trivial Born approximation result, while in the range $|\varepsilon|\lesssim\varepsilon_{\rm min}^{\rm (B)}$  the non-perturbative problem was solved due to the possibility of reduction to evaluation of the exact strictly one-dimensional density of states. The latter was studied already long ago \cite{FrishLloyd1960} with the use of gaussian character of random potential at  $n\gg n_c$. The results for repulsing and attracting impurities are identical; the cases of a tube and a strip are very similar: only some numerical coefficients differ.

For $n\ll n_c$ we have found the resistivity  in the energy range $|\varepsilon|\gg\varepsilon_{\rm min}^{\rm (nB)}$. The obtained result is a perturbative one only for $|\varepsilon|\gg\varepsilon_{\rm nB}$, while for $\varepsilon_{\rm min}^{\rm (nB)}\ll|\varepsilon|\lesssim\varepsilon_{\rm nB}$ strong non-Born renormalization of scattering was taken into account. It was feasible because in this range of energies only the single-impurity renormalizations are relevant, the multi-impurity ones come into play only at 
$|\varepsilon|\lesssim\varepsilon_{\rm min}^{\rm (nB)}$. The results for tube and for strip differ from each other quite substantially, because of the position-dependence of scattering amplitudes of different impurities in the strip case. Also the difference between cases of repulsing and attracting impurities is dramatic. 

The range of parameters $n\ll n_c$, $|\varepsilon|\lesssim\varepsilon_{\rm min}^{\rm (nB)}$, where the multi-impurity renormalization of scattering dominates, was not studied in the present paper. However, it seems to be important because just in this range the resistivity reaches its minimum. An approach to this problem will be discussed in a separate publication.

This work was supported by Basic Research Program of The Higher School of Economics. The research of N. Peshcherenko was also supported by the Foundation for Advancement of Theoretical Physics and Mathematics ``Basis'' under grant 20-1-5-150-1.

\appendix

\section{Evaluation of functions $F_1(\epsilon)$ and $\tilde{F}_1(\epsilon)$.\label{Evaluation of function1}}

To perform the integration in \eqref{res_repuls}, \eqref{rho_attract}, \eqref{res_repulsr},  we note that $F_1,F_2,\tilde{F}_1$ can be presented as contour integrals around the cut between $t=0$ and $t=1$ on the Riemannian surface of complex  $t$. 
In particular, we have
\begin{align}
 F_1(\epsilon)=\frac{2\sqrt{\epsilon}}{\pi}\int_0^1\frac{t^{1/2}(1-t)^{-1/2}dt}{(\epsilon+4t^2)}=\nonumber\\=\frac{\sqrt{\epsilon}}{\pi}\oint_{{\rm cut}\;[0,1]}\frac{t^{1/2}(1-t)^{-1/2}dt}{(\epsilon+4t^2)}=\nonumber\\=\frac{\sqrt{\epsilon}}{2}{\rm Re}\,\left\{\frac{i}{\sqrt{t_0(1-t_0)}}\right\}=\left(\frac{\sqrt{1+4/\epsilon}-1}{2(1+4/\epsilon)}\right)^{1/2}
 \label{deni2pp4sa0}
\end{align}
where
\begin{align}
t_0=i\sqrt{\epsilon}/2 \label{deni2pp4sal}
\end{align}
is the position of the integrand's pole in the upper half-plane of complex $t$. This pole is responsible for the virtual bound state, causing the resonant enhancement of slow electrons, mentioned in Section \ref{Paradox}. 

In a similar manner
\begin{align}
 \tilde{F}_1(\epsilon)=\frac{2\sqrt{\epsilon}}{\pi}\oint_{{\rm cut}\;[0,1]}\frac{t^{3/2}(1-t)^{-1/2}dt}{(\epsilon+4t^2)}=\nonumber\\=\sqrt{\epsilon}{\rm Re}\,\left\{i\sqrt{\frac{t_0}{1-t_0}}\right\}+\sqrt{\epsilon}=\nonumber\\=\sqrt{\epsilon}\left\{1-\sqrt{\frac{1+\sqrt{1+4/\epsilon}}{2(1+4/\epsilon)}}\right\}\approx\begin{cases} \sqrt{\epsilon},\quad \epsilon\ll1\\
\frac{3}{2\sqrt{\epsilon}},\quad \epsilon\gg1.
\end{cases}
 \label{deni2pp4sa0vb}
\end{align}

Note that the results \eqref{deni2pp4sa0}, \eqref{deni2pp4sa0vb} are valid  for both repulsing and attracting impurities (i.e., for either signs of $\lambda$), the only necessary requirement being $\epsilon>0$.

\section{Evaluation of function $F_2(\epsilon,\lambda)$.\label{Evaluation of function2}}

For $\epsilon<0$ the pole of the integrand in the leading approximation in $\lambda$ is 
\begin{align}
t_0=-{\rm sign}\,\lambda\sqrt{-\epsilon}/2 \label{deni2pp4sap}
\end{align}
and lies  on the real axis. In particular, it may occur directly on the cut, provided $\lambda<0$ and $-4<\epsilon<0$. It would lead to formal divergency of $F_2(\epsilon,\lambda\to -0)$. Just in order  to remove this divergency we have taken into account
 the additional term in the denominator of \eqref{rho_attract} which leads to a shift of $z_0$ away from the real axis:
\begin{align}
t_0\to \sqrt{-\epsilon}(1\pm i|\lambda|)/2. \label{deni2pp4sap1}
\end{align}
Note that initially the pole $t_0$ was the second order one. Taking into account additional small term leads to splitting it into two first order poles. 
 
\subsection{Repulsing impurities.\label{Evaluation of function2r}} 

For $\lambda>0$ the imaginary part of $t_0$  may be neglected, so we have $t_0=-\sqrt{-\epsilon}/2<0$ and
\begin{align}
 F_2(\epsilon,\lambda)\approx F_2(\epsilon,\lambda\to+0)=\nonumber\\=\frac{1}{2\pi}\oint_{{\rm cut}\;[0,1]}\frac{t^{-1/2}(1-t)^{-1/2}dt}{\left(1+2t/\sqrt{-\epsilon}\right)^2}=\nonumber\\=\frac{-i\epsilon}{4}\left(\frac{d}{dt}\sqrt{\frac{1}{t(1-t)}}\right)_{t_0}=\frac{-i\epsilon}{8}\left(\frac{2t_0-1}{\sqrt{t_0^3(1-t_0)^3}}\right)=\nonumber\\=(-\epsilon)^{1/4} \left(1+\sqrt{-\epsilon}\right)\left(2+\sqrt{-\epsilon}\right)^{-3/2}.
 \label{tauresnegeval}
\end{align}

\subsection{Attracting impurities, outside the impurity band, $\epsilon<-4$.\label{Evaluation of function2a}}

For $\epsilon<-4$ the pole \eqref{deni2pp4sap} lies outside the cut even for $\lambda<0$.
Thus, for $\lambda<0$ we can simply put $\lambda\to-0$ and obtain the relevant result by a substitution $\sqrt{-\epsilon}\to-\sqrt{-\epsilon}$ in \eqref{tauresnegeval}:
\begin{align}
F_2(\epsilon,\lambda)\approx F_2(\epsilon,\lambda\to-0)=\nonumber\\=(-\epsilon)^{1/4} \left(\sqrt{-\epsilon}-1\right)\left(\sqrt{-\epsilon}-2\right)^{-3/2}.
 \label{deni2pp4s100i}
\end{align}

\subsection{Attracting impurities, within the impurity band, $-4<\epsilon<0$.\label{Evaluation of function2a1}}

Here, to obtain a finite result we have to take into account the additional term in the denominator of \eqref{rho_attract} or, equivalently, the imaginary part of $t_0$, so that for $\lambda<0$ and $-4<\varepsilon<0$ the function $F_2(\epsilon,\lambda)$ is a function of two dimensionless variables: one cannot put $\lambda\to -0$, but has to keep it finite. Then \eqref{rho_attract} can be rewritten in a form
\begin{align}
F_2(\epsilon,\lambda)=\frac{1}{\pi}\int_0^1\frac{t^{-1/2}(1-t)^{-1/2}dt}{\left(1-2t/\sqrt{-\epsilon}\right)^2-4t^2\lambda^2/\epsilon}\approx\nonumber\\\approx\frac{1}{\pi}\frac{-\epsilon}{4}\int_0^1\frac{t^{-1/2}(1-t)^{-1/2}dt}{\left(\sqrt{-\epsilon}/2-t\right)^2-\epsilon|\lambda|^2/4},\label{eett9}
\end{align}
where the second term in the denominator was substituted by its value at resonance: $-4t^2\lambda^2/\epsilon\to\lambda^2$.
Since $|\lambda|\ll 1$ it is possible to write
\begin{align}
\frac{1}{\left(\sqrt{-\epsilon}/2-t\right)^2-\epsilon\lambda^2/4}\approx\frac{2\pi}{\sqrt{-\epsilon}|\lambda|}\delta \left(\sqrt{-\epsilon}/2-t\right),\label{deni2pp4po3pp}
\end{align}
and get
\begin{align}
F_2(\epsilon,\lambda)\approx\frac{1}{|\lambda|}\left(\frac{\sqrt{-\epsilon}}{2-\sqrt{-\epsilon}}\right)^{1/2}.\label{leeft1}
\end{align}

\subsection{Attracting impurities, near the edge of the impurity band, $|\epsilon+4|\ll 1$.\label{Evaluation of function2a2}}

Both the results \eqref{leeft1} and \eqref{deni2pp4s100i} formally diverge as $\epsilon\to-4$, so they are apparently not applicable in the narrow vicinity of $\epsilon=-4$, namely, for $|\epsilon+4|\lesssim 8|\lambda|$. In this range we should write
\begin{align}
F_2(\epsilon,\lambda)\approx\frac{1}{\pi}\int_0^1\frac{t^{-1/2}(1-t)^{-1/2}dt}{[1-t-(\epsilon+4)/8]^2+\lambda^2}\approx\nonumber\\\approx\frac{1}{|\lambda|^{3/2}}\Phi\left(-\frac{\epsilon+4}{8|\lambda|}\right),
\label{deni2pp4po33dnn}
\end{align}
where
\begin{align}
\Phi(a)=\frac{1}{\pi}\int_\infty^{-\infty}\frac{d \phi}{(\phi^2+a)^2+1}=-
{\rm Im}\;\left\{\frac{1}{\sqrt{a+i}}\right\}=\nonumber\\=
\left(\frac{\sqrt{a^2+1}-a}{2(a^2+1)}\right)^{1/2}.
\label{Phifunc}
\end{align}

\end{document}